\documentclass[a4paper,11pt]{article}
\pdfoutput=1 

\usepackage{jheppub} 

\usepackage[T1]{fontenc} 
\usepackage{graphicx}
\usepackage{amsmath}
\usepackage{amssymb}
\usepackage{array}
\usepackage{slashed}
\usepackage{comment}
\usepackage{datetime}
\usepackage{soul}
\usepackage{multirow}
\usepackage{pifont}

\usepackage{float}
\usepackage{bm}
\usepackage{bbm,multicol}

\usepackage{subfigure}
\usepackage{dsfont}

\usepackage{marginnote}

\newcommand{\met}{\ensuremath{{\not\mathrel{E}}_T}}
\newcommand{\cmark}{\ding{51}}%
\newcommand{\xmark}{\ding{55}}%

\def\h{h^0}
\def\H{H^0}
\def\A{A}
\newcommand{\sba}{\ensuremath{\sin(\beta-\alpha)}}
\newcommand{\cba}{\ensuremath{\cos(\beta-\alpha)}}

 \title{Exotic Decays Of A Heavy Neutral Higgs Through HZ/AZ Channel}
\author{Baradhwaj Coleppa, Felix Kling, Shufang Su}
\emailAdd{baradhwa@email.arizona.edu, kling@email.arizona.edu, shufang@email.arizona.edu}
\affiliation{1118 E. 4th st., P.O. Box 210081,\\Department of Physics,
University of Arizona,
Tucson, AZ 85721, USA}

\abstract{Models of electroweak symmetry breaking with extended Higgs sectors are theoretically well motivated. In this study, we focus on the Two Higgs Doublet Model with a low energy spectrum containing  scalars $H$ and a pseudoscalar $\A$. We study the decays  $\A\rightarrow H Z$ or $H\rightarrow \A Z$, which could reach sizable branching fractions in certain parameter regions.   With detailed collider analysis, we obtain     model independent exclusion bounds as well as  discovery reach at the 14 TeV LHC  for  the process:  $gg\rightarrow \A/H \rightarrow H Z/\A Z$,  looking at final states $bb\ell\ell$, $\tau\tau\ell\ell$ and $ZZZ(4\ell+2j)$ for $\ell =e$, $\mu$.   We further interpret these bounds   in the context of the Type II Two Higgs Doublet Model, considering three different classes of processes:  $\A\to\h Z$, $\A\to\H Z$, and $\H\to\A Z$, in which $\h$ and $\H$ are the light and heavy CP-even Higgses  respectively.  For 100 ${\rm fb}^{-1}$ integrated luminosity at the 14 TeV LHC, we find that  for parent particle mass around  300 $-$ 400 GeV, $\A\to\h Z$ has the greatest reach when $\H$ is interpreted as the 126 GeV Higgs: most regions in the  $\tan\beta$ versus $\sba$ plane can be excluded and a significant fraction at small and large $\tan\beta$ can be covered by discovery.   For 126 GeV $\h$,   only relatively small $\tan\beta \lesssim 10$ (5) can be reached by exclusion (discovery) while a wide range of $\sba$ is accessible.  For $\A\to\H Z$, the reach is typically restricted to $\sba\sim \pm 1$ with $\tan\beta \lesssim 10$ in $bb\ell\ell$ and $\tau\tau\ell\ell$ channels.   The $ZZZ (4 \ell 2j)$ channel, on the other hand, covers a  wide range of $0.3 < |\sba| < 1$ for $\tan\beta \lesssim 4$.  $\H\to\A Z$ typically  favors negative values of $\sba$, with exclusion/discovery reach possibly extending to all values of $\tan\beta$.  A study of exotic decays of extra Higgses appearing in extensions of the Standard Model would extend the reach at the LHC and provides nice  complementarity to 
conventional Higgs search channels.  }

\begin{document}

\maketitle
\flushbottom
\newpage

\section{Introduction}
\label{sec:intro}

The greatest experimental triumph of the Large Hadron Collider (LHC) till date is the discovery of a scalar resonance at 126 GeV with properties consistent with that of the Standard Model (SM) Higgs \cite{Aad:2012tfa, ATLAS:2013sla, Chatrchyan:2012ufa,CMS:yva}.  The mass of this particle along with its spin \cite{ATLAS:2013sla,CMS:yva,Aad:2013xqa} has now been established, and a complete characterization of all its possible decay modes is underway. At the same time, from the theoretical front, we have now known for a while that the SM, though in excellent agreement with experiments, has to be supplanted with other dynamics if it is to explain many puzzles facing particle physics today, viz., the hierarchy problem, neutrino masses, and the nature of dark matter, to name a few. Many beyond the SM scenarios are constructed to explain one or many of these puzzles, and are becoming more constrained by the Higgs observation   at the LHC. This is particularly true for theories constructed with an extended Higgs   sector.  Well known examples  are the Minimal Supersymmetric Standard Model (MSSM) \cite{Nilles:1983ge,Haber:1984rc,Barbieri:1987xf}, Next to Minimal Supersymmetric Standard Model (NMSSM) \cite{Ellis:1988er,Drees:1988fc} and Two Higgs Doublet Models (2HDM) \cite{Branco:2011iw,type1,hallwise,type2}.  In addition to the SM-like Higgs boson in these models, the low energy spectrum includes other CP-even Higgses, CP-odd Higgses, as well as charged ones.  

Models with an extended Higgs sector hold a lot of phenomenological interest. The discovery of   extra Higgses  would be an unambiguous evidence for new physics beyond the SM.  Other than the decay of these extra Higgses into the SM final states $\gamma\gamma$, $ZZ$,    $WW$,   $bb$ and $\tau\tau$, which have been the focus of the current Higgs searches, the decay of heavy Higgses into light Higgses, or Higgs plus gauge boson final states could also be sizable.  Such decays are particularly relevant as the 126 GeV resonance could show up as a decay of a heavier state, opening up the interesting possibility of using the SM-like Higgs to discover its heavier counterparts.   It is thus timely to study these exotic Higgs decay channels and fully explore the experimental discovery potential for the enlarged Higgs sector.  

In this paper, we focus on  the decays $H\rightarrow \A Z$ or $\A\rightarrow H Z$, with $H$ and $\A$ referring to generic CP-even and CP-odd Higgs, respectively\footnote{Note that we use $\h$ and $\H$ to refer to the lighter or the heavier CP-even Higgs for models with two CP-even Higgs bosons.  When there is no need to specify, we use $H$ to refer to the CP-even Higgses.}. We consider leptonic decays of the $Z$, with the $\A/H$ in the final states decaying to either a pair of fermions ($bb$ or $\tau\tau$) or $ZZ$ and  explore the exclusion bounds as well as discovery reach at the LHC for various combinations of $(m_{\A}, m_{H})$.

In the 2HDM or NMSSM, both decays  $H_i\rightarrow \A_j Z$ and $\A_i\rightarrow H_j Z$ could appear with large branching fractions as shown in~\cite{Craig:2012vn, Chiang:2013ixa, Christensen:2013dra, Grinstein:2013npa}.   Ref.~\cite{Lewis:2013fua} also argued that $A\rightarrow \h Z$ could have a sizable branching fraction in the  low $\tan\beta$ region   of the MSSM with the light CP-even $\h$ being SM-like.   A brief Snowmass study of $\A/H\rightarrow H Z/\A Z$ with $bb\ell\ell$ final state can be found in Ref.~\cite{Coleppa:2013xfa}.    Another  Snowmass study of heavy Higgses  \cite{Brownson:2013lka}  explored sensitivities in the $\H\to ZZ \to 4\ell$ and $\A\to Z\h \to bb\ell\ell$,  $\tau\tau\ell\ell$ channels at the 14 TeV and 33 TeV LHC, focusing on the case with $\h$ being the 126 GeV Higgs.   In our study, we consider   a variety   of   daughter Higgs masses in $bb\ell\ell$ and $\tau\tau\ell\ell$ channels,  and analyze $\A \rightarrow \H Z \rightarrow ZZZ$ in addition.   We also interpret  the  
search  results   in the context of the Type II 2HDM.

The paper is organized as follows.  In Sec.~\ref{sec:scenarios}, we present a brief overview of models and parameter regions  where the channels under consideration   can be significant.  In Sec.~\ref{sec:limits}, we summarize the current experimental search limits on heavy Higgses.    In Sec.~\ref{sec:bb},  we present the details of the  analysis of the $HZ/AZ$ with the $bb\ell\ell$ final states.  We also show   model-independent results of 95\% C.L. exclusion as well as 5$\sigma$ discovery limits for $\sigma \times BR(gg \rightarrow A/H \rightarrow H Z/A Z \rightarrow bb \ell \ell)$ at the 14 TeV LHC with 100, 300 and 1000 ${\rm fb}^{-1}$ integrated luminosity.  In Secs.~\ref{sec:tautau} and \ref{sec:ZZZ}, we present the analysis for the  $\tau\tau\ell\ell$ and $ZZZ$ final states, respectively.   In Sec.~\ref{sec:implications}, we study the implications of the collider search limits on the parameter regions of the Type II 2HDM.   We conclude in Sec.~\ref{sec:conclusions}.

  \section{Scenarios with large $H\rightarrow \A Z$ or  $\A\rightarrow H Z$}
 \label{sec:scenarios}
  
In the 2HDM,  we introduce two ${\rm SU}(2)$ doublets   $\Phi_{i}$,  $i=1,2$:
 \begin{equation}
\Phi_{i}=\begin{pmatrix} 
  \phi_i^{+}    \\ 
  (v_i+\phi^{0}_i+iG_i)/\sqrt{2}  
\end{pmatrix},
\label{eq:doublet}
\end{equation} 
where $v_1$ and $v_2$ are the vacuum expectation values  of the neutral components which satisfy the relation: $\sqrt{v_1^2+v_2^2}=$ 246 GeV after electroweak symmetry breaking.  Assuming a discrete ${\cal Z}_2$ symmetry imposed on the Lagrangian,  we are left with six free parameters, which can be chosen as    four Higgs masses ($m_h$, $m_H$, $m_A$, $m_{H^{\pm}}$), the mixing angle $ \alpha$ between the two CP-even Higgses, and the ratio of the two vacuum expectation values, $\tan\beta=v_2/v_1$.   In the case in which a soft breaking of the ${\cal Z}_2$ symmetry is allowed, there is an additional parameter   $m_{12}^2$.

The mass eigenstates contain a pair of CP-even Higgses: $\h$, $\H$, one CP-odd Higgs, $\A$ and  a pair of charged Higgses $H^\pm$\footnote{For more details about the model, see Ref.~\cite{Branco:2011iw}.}:
\begin{equation}
\left(\begin{array}{c}
\H\\ \h
\end{array}
\right)
=\left(
\begin{array}{cc}
\cos\alpha &\sin\alpha\\
-\sin\alpha&\cos\alpha
\end{array}
\right)  \left(
\begin{array}{c}
\phi_1^0\\\phi_2^0
\end{array}
\right),\ \ \ 
\begin{array}{c}
 \A \\H^\pm
 \end{array}
 \begin{array}{l}
 =  -G_1\sin\beta+G_2\cos\beta\\
 =-\phi_1^{\pm}\sin\beta+\phi_2^{\pm} \cos\beta
 \end{array}.
 \label{eq:mass}
 \end{equation}

Two types of couplings that are of particular interest are $Z\A\H/\h$ couplings and $\H/\h VV$ couplings, with $V$ being   the SM gauge bosons $W^\pm$ and $Z$.   Both are determined by the gauge coupling structure and the mixing angles. The couplings for $Z\A\H$ and $Z\A\h$ are \cite{Gunion:1989we}: 
\begin{equation}
 g_{Z\A\H}=-\frac{g\sba}{2\cos\theta_w}( p_{\H} -p_A)_\mu, \ \ \ 
 g_{Z\A\h}=\frac{g\cba}{2\cos\theta_w}(p_{\h}-p_A)_\mu,
 \label{eq:haz-coup}
 \end{equation}
 with $g$ being the ${\rm SU}(2)$ coupling, $\theta_w$ being the Weinberg angle and $p_\mu$ being the incoming momentum of the corresponding   particle.
 
The $\H VV$ and $\h VV$ couplings are:
 \begin{equation}
 g_{\H VV}=\frac{m_V^2}{v}\cba,\ \ \ 
 g_{\h VV}=\frac{m_V^2}{v}\sba.
  \label{eq:hvv-coup}
 \end{equation}
 
Note that $\A$ always couples to the non-SM-like Higgs more strongly.   If we demand $\h$ ($\H$) to be SM-like, then $|\sba|\sim 1$ ($|\cba|\sim 1$) is preferred, and the $Z\A\H$ ($Z\A\h$) coupling is unsuppressed.  Therefore, in the $\h$-126 case,  $\A$ is more likely to decay to $\H Z$ than $\h Z$, unless the former decay is kinematically suppressed.  $\H \rightarrow \A Z$ could also be dominant once it is kinematically open. Particularly for a heavy $\H$, as we will demonstrate later in  Sec.~\ref{sec:implications},   $\H \rightarrow \A Z$ can have a  large branching fraction in the $\sba=\pm$1 regions.  On the contrary,  for $\H$ being SM-like with  $|\cba|\sim 1$, $\A \rightarrow \h Z$ dominates over $\H Z$ channel.  For very light $m_A$, $\h \rightarrow \A Z$ could also open.   The detectability of this channel, however,  is challenging given the soft or collinear final decay products from a light $A$.    Therefore, for our discussion below, we will focus on the cases    $\A \rightarrow \h Z,\ \H Z$ 
and $\H \rightarrow \A Z$ only. 

In the generic 2HDM, there are no mass relations between the pseudoscalar and the scalar states.  Thus, the decays $\A \rightarrow \h Z,\ \H Z$ and $\H \rightarrow \A Z$  can happen in different regions of parameter spaces. It was shown in  Ref.~\cite{Coleppa:2013dya}  that in  the Type II 2HDM with ${\cal Z}_2$ symmetry, imposing all experimental and theoretical constraints still leaves sizable regions in the parameter space.  In those parameter spaces,  such exotic decays can have unsuppressed decay branching fractions.   It was also pointed out in Ref.~\cite{Branco:2011iw} that in the Type I 2HDM, for $\cos^2(\alpha-\beta)> 1/2$, the decay $\h\to\A Z$ will actually dominate the $WW$ decay for a light $A$.    Results obtained in this study can also be applied to the CP-violating 2HDM in which $H_i \rightarrow H_j Z$ could be sizable with $H_{i,j}$ being mixtures of CP-even and CP-odd states.  Appropriate rescaling of the production cross sections and decay branching fractions is needed to recast the results.

The Higgs sector in the MSSM is more restricted, given that the quartic Higgs couplings are fixed by the gauge couplings and the tree-level Higgs mass matrix only depends on $m_A$ and $\tan\beta$.  In the usual decoupling region with large $m_A$,  the light CP-even Higgs $h^0$ is SM-like while the  other  Higgses are almost degenerate: $m_{\H} \sim m_{A} \sim m_{H^\pm}$.  Thus,  $\A\to Z\H$ or $\H\to Z\A$ is not allowed   kinematically.  $\A\to Z\h$ is typically suppressed by the small coupling:  $\cba\sim 0$,   and is only relevant for small $\tan\beta$.   In the NMSSM, the Higgs  sector of MSSM is enlarged to include an additional singlet. It was shown in Ref.~\cite{Christensen:2013dra} that there are regions of parameter space where the decay $\A_i\to  H_j Z$ can be significant.

  \section{Current experimental limits}
 \label{sec:limits}
 
 Searches for the non-SM like Higgses, mainly in the    $bb$, $\mu\mu$, $\tau\tau$ or $WW/ZZ$ channels have been performed both by ATLAS and CMS. No evidence for a neutral non-SM like Higgs was found.

Searches for the neutral Higgs bosons $\Phi$ of the MSSM   in the process $pp\to \Phi \to \mu^+\mu^-/\tau^+\tau^-$ have been performed by the ATLAS \cite{Aad:2012cfr},  and in the $\tau^+\tau^-$channel  at CMS \cite{CMS-tautau}.  Limits in the $\mu\mu$ channel are much weaker given the extremely small branching fraction in the MSSM. The production mechanisms considered were both gluon fusion and $bb$ associated production, and the exclusion results were reported for the MSSM $m_h^{\rm{max}}$ scenario.  The ATLAS study was performed at $\sqrt{s}=7$ TeV with 4.7 - 4.8 fb$^{-1}$ integrated luminosity looking at three different possible $\tau\tau$ final states, $\tau_e\tau_{\mu}$, $\tau_{\rm{lep}}\tau_{\rm{had}}$, and $\tau_{\rm{had}}\tau_{\rm{had}}$.  The ATLAS search rules out a fairly sizable portion of the MSSM parameter space, extending from about $\tan\beta$ of 10 for $m_{\A} \sim$ 130 GeV, to $\tan\beta \approx 60$ for $m_{\A}$= 500 GeV. The corresponding exclusion in $\sigma_{\Phi}\times\rm{BR}(\Phi\to\tau\tau)$ extends from roughly 40 pb to 0.3 pb  in that mass range.   The CMS study was performed with 19.7 fb$^{-1}$ integrated luminosity at 8 TeV and 4.9 fb$^{-1}$ at 7 TeV in the $\tau_e\tau_{\mu}$, $\tau_{\mu}\tau_{\mu}$, $\tau_{\rm{lep}}\tau_{\rm{had}}$, and $\tau_{\rm{had}}\tau_{\rm{had}}$ final states. The search excludes roughly between $\tan\beta$ of 4 for $m_{\A}$= 140 GeV and  $\tan\beta \approx  60$ for $m_{\A}$= 1000 GeV. The corresponding exclusion in $\sigma_{\Phi}\times\rm{BR}(\Phi\to\tau\tau)$ extends from roughly 2 pb to 13 fb in that mass range.

\begin{figure}[h!]
\begin{center}
	\includegraphics[scale=1]{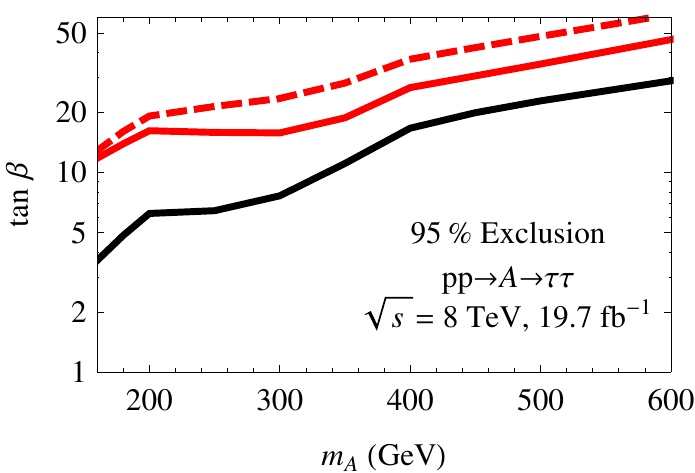}
	\includegraphics[scale=1]{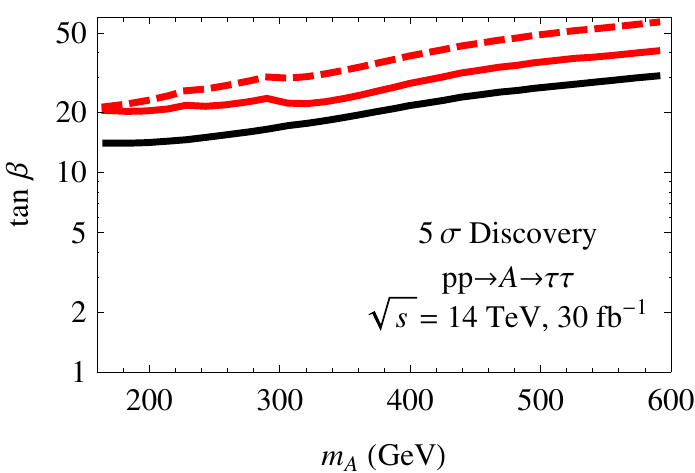}
  \caption{The reach of $pp \rightarrow A \rightarrow \tau\tau$ in  $m_A-\tan\beta$ parameter space of the Type II 2HDM.  Left panel shows the current 95\% C.L. exclusion limits from  CMS ~\cite{CMS-tautau} with 19.7 ${\rm fb^{-1}}$ data collected at the $\sqrt{s}=8$ TeV LHC.  Right panel shows the projected 5$\sigma$ discovery reach at the 14 TeV LHC with 30 fb$^{-1}$ luminosity~\cite{atlas-tautau-future}.  In both plots, the solid black curves correspond to the limits in the MSSM, when $m_A \approx m_{\H}$ with both $\A$ and $\H$ contributing to the signal.  The solid red curves correspond to the limits in the type II 2HDM, when only contribution from $\A$ is included and $\H$ is decoupled.   Also shown in the red dashed curves are  the reduced $\tau\tau$ channel limits when $\A\to\h Z$ is open with the parameter choice of $\sba=0$, $m_{\h}=50$ GeV and $m_{\H}=126$ GeV. 
  }
\label{fig:atlastautau}
\end{center}
\end{figure}

In  Fig.~\ref{fig:atlastautau}, we recast the current 95\% C.L. limit   of $pp\to \Phi \to \tau^+\tau^-$ in the $(m_A,\tan\beta)$ parameter space of the Type II 2HDM~\cite{CMS-tautau} (left panel) and   the projected 5$\sigma$ reach at the 14 TeV LHC with 30 ${\rm fb}^{-1}$ luminosity~\cite{atlas-tautau-future} (right panel).     In both plots, the solid black curves correspond to the limits in the MSSM, when $m_A \approx m_{\H}$ with both $\A$ and $\H$ contributing to the signal.  The solid red curves correspond to the limits in the type II 2HDM, when only contribution from $\A$ is included and $\H$ is decoupled.   The reach is considerably weaker: the current exclusion is about $\tan\beta\sim 12 $ at $m_A= 160$ GeV, and $\tan\beta \sim 46$ for $m_A = 600$ GeV. At the 14 TeV LHC with 30 ${\rm fb}^{-1}$ luminosity,  the $5\sigma$ reach extends beyond the current exclusion for large $m_A$.   Dashed lines indicate the reduced reach in the $\tau\tau$ channel once $A\rightarrow \h Z$ mode opens, for a benchmark point of $\sba=0$, $m_{\h}=50$ GeV and $m_{\H}=126$ GeV.  

Searches with $bb$ final states have also been performed for the MSSM Higgs in the associated production $pp\to b \Phi +X$.   The CMS search, done with 2.7 $-$ 4.8 fb$^{-1}$ of data at $\sqrt{s}=7$ TeV excludes $\tan\beta$ values between 18 and 42 in the mass range 90 GeV $<m_{\A}<$ 350 GeV \cite{CMS-bb}.

The ATLAS collaboration has also looked for the heavier CP-even Higgs in the Type I and Type II 2HDM, assuming the lighter CP-even   Higgs is the discovered 126 GeV boson \cite{atlas-WW2HDM}. The study was performed with 13 fb$^{-1}$ integrated luminosity at 8 TeV and considered both gluon fusion and vector boson fusion production.    Searches in the process   $\H \to WW \to e\mu\nu_{e}\nu_{\mu}$ exclude a significant region of the $m_{\H}-\cos\alpha$ parameter space in the mass range 135 GeV$< m_{\H} <$ 200 GeV for the Type II 2HDM.  The excluded region shrinks for higher $\tan\beta$ due to the reduced  branching ratio to $WW$. This would serve as a useful constraint if we were to look at decays of the relatively light $\H$ to light $\A$'s. In this paper, we consider values of $m_H$  outside this mass range so this constraint does not apply.     

The CMS collaboration has also searched for the heavier CP-even Higgs $\H$ and a heavy CP-odd Higgs $\A$ in 2HDM via the processes $gg \rightarrow \A \to \h Z$ and $gg \rightarrow\H \to \h \h$, assuming the lighter Higgs $\h$ is the discovered 126 GeV boson \cite{CMS-HZ}. The study was performed with 19.5 fb$^{-1}$ integrated luminosity at 8 TeV. Various possible decays of the SM-Higgs were taken into account.    Assuming  SM branching ratios   for $\h$, this study gives an upper bound on $\sigma \times { \rm{BR}} (\A \to \h Z)$ of roughly 1.5 pb for   $m_{\A} $ between 260 and 360 GeV and  $\sigma \times {\rm{BR}}(\H \to \h \h)$ between 8 pb and 6 pb for masses $m_{\H}$ between 260 GeV and 360 GeV. The corresponding excluded parameter space for the Type II 2HDM in the $\tan \beta -\cos(\beta - \alpha)$ plane was also analyzed.  In the analysis presented in this paper,  we do not necessarily require that the daughter Higgs in $A \to H Z$ to be the SM-like Higgs or  have SM-like branching ratios.  Furthermore 
we also analyze the process $H \to A Z$ for light $A$ and its implication in the Type II 2HDM. 

 
  \section{Collider analysis}
  
In this section, we will present model \emph{independent} limits on the $\sigma\times\rm{BR}$ for both 95\% C.L.  exclusion and 5$\sigma$ discovery for $A/H \rightarrow HZ/AZ$ in the various final states of $bb\ell\ell$, $\tau\tau\ell\ell$ and $ZZZ(4\ell2j)$.     In this study we focus on the leptonic decay of the $Z$, which allows   precise mass reconstruction and
 suppresses the background sufficiently. Other decay modes of the $Z$, for example $Z \to \tau\tau$,
might be useful in studying this channel as well.   In  the discussion of the analyses and results below, we use the decay $A \rightarrow HZ$ for $m_A > m_H + m_Z$ as an illustration.    Since we do not make use of angular correlations, the bounds obtained for $\A\rightarrow H Z$ apply to  $H\rightarrow \A Z$  as well with the values of $m_{\A}$ and $m_{H}$ switched.       
   
 \subsection{$\A/H\to HZ /\A Z\rightarrow bb\ell\ell$}
 \label{sec:bb}
We start our analysis by looking at the channel $\A/ H\to HZ/\A Z\rightarrow bb\ell\ell$ for $\ell=e,\mu$,  focusing only on the gluon fusion production channels.   We use $H$ to refer to either the light or the heavy CP-even Higgs.  Since the only allowed couplings are of the type $H-\A-Z$,  if  the parent particle is a scalar $H$, the daughter particle is necessarily a pseudoscalar $\A$ and vice versa.   

The dominant SM backgrounds for $bb \ell \ell $ final states are $Z/\gamma^*bb$ with leptonic $Z/\gamma^*$ decay, $t\bar{t}$ with leptonically decaying top quarks, $ZZ \rightarrow bb\ell \ell $, and $H_{\rm SM}Z$~\cite{Cordero:2009kv, Kidonakis:2012db,Campbell:2011bn,Dittmaier:2011ti}.  We have ignored the subdominant backgrounds from $WZ$, $WW$, $H_{\rm SM} \rightarrow ZZ$, $Wbb$, Multijet QCD Background, $Zjj$, $Z\ell\ell$ as well as  $tWb$.   These backgrounds either have small production cross sections, or can be sufficiently suppressed by the cuts imposed. We have included  $H_{\rm SM}Z$ here even if the cross section is very small because it  has the same final state as the process under consideration, especially for the $\A \rightarrow H_{\rm SM}Z$ case.  The total cross sections for these backgrounds can be found in Table \ref{tab:bbcuts}.

We use Madgraph 5/MadEvent v1.5.11 \cite{Alwall:2011uj} to generate our signal and background events.  These events are passed to Pythia v2.1.21 \cite{PYTHIA}  to simulate initial and final state radiation,  showering and hadronization. The events are further passed through Delphes 3.09 \cite{deFavereau:2013fsa} with the Snowmass combined LHC detector card \cite{snowmassdetector} to simulate detector effects.  

For the signal process, we generated event samples at the 14 TeV LHC for $gg \rightarrow \A \rightarrow H Z$  with the daughter particle mass fixed at 50, 126, and 200   GeV while varying the parent particle mass  in the range of 150 $-$ 600 GeV.  We applied the following cuts to identify the signal from the backgrounds\footnote{Requiring the missing transverse energy  to be small would potentially greatly reduce the $t\bar{t}$ background.  However, including  pile-up effects introduces $\met $ in the signal events, which renders the cut inefficient.     We thank Meenakshi Narain and John Stupak for pointing this out to us. }:   
\begin{enumerate}
	\item \textbf{Two isolated leptons, two tagged $b$'s:} 
 \begin{equation}
n_{\ell}=2,  \ n_{b} =2,\ {\rm with}\ |\eta_{\ell,b} |<2.5,\ p_{T,\ell} > 10\ {\rm GeV},\ p_{T,b}>15\ {\rm GeV}.
\label{cut_1}
\end{equation}
 For jet reconstruction, the anti-$k_T$ jet algorithm with $R= 0.5$ is used. 

	\item \textbf{Lepton trigger}   \cite{citation_trigger}{\bf :}
\begin{equation}
 p_{T,\ell_1}>30 \text{ GeV}\ \text{or} \ p_{T,\ell_1}>20 \text{ GeV},\ p_{T,\ell_2}>10 \text{ GeV}.
\label{cut_2}
\end{equation}

	\item \textbf{Dilepton mass $m_{\ell \ell}$:} We require the dilepton mass to be in the $Z$-mass window:
\begin{equation}
80 \text{ GeV} <m_{\ell \ell}<  100 \text{ GeV}.
\label{cut_3}
\end{equation}		
	
	\item \textbf{$m_{bb}$ versus $m_{bb\ell \ell}$:} We require the dijet mass $m_{bb}$ to be close to the daughter-Higgs mass $m_{H}$ and the mass $m_{bb\ell\ell}$ to be close to the parent-Higgs mass $m_{\A}$. These two invariant masses are correlated, i.e., if we underestimate $m_{bb}$ we also underestimate $m_{bb\ell \ell}$. To take this into account we apply a two-dimensional cut: 
\begin{equation}
\begin{aligned}
(0.95 - w_{bb})\times m_{H} < \; &m_{bb} < (0.95 + w_{bb})\times m_{H} \text{ with } w_{bb}= 0.15, \\
\frac{m_Z + m_{H}}{m_{\A}}\times(m_{bb\ell\ell}-  m_{\A} - w_{bb\ell \ell} )< m_{bb} & -m_{H} < \frac{m_Z + m_{H}}{m_{\A}}\times(m_{bb\ell\ell } -m_{\A} + w_{bb\ell \ell} ), 
\end{aligned}
\label{cut_4}
\end{equation}	
where $w_{bb} \times m_H$ is the width of the dijet mass window. Note that the slightly shifted reconstructed Higgs mass $m_{bb}$ (0.95 $m_{H}$ instead of $m_{H}$) is due to the reconstruction of the $b$-jet with a small size of $R=0.5$. The second condition describes two lines going through the points $(m_{\A} \pm w_{bb\ell \ell }, m_{H})$ with slope $(m_Z + m_{H})/m_{\A}$. We choose a width for the $m_{bb\ell \ell }$ peak of $w_{bb\ell \ell } = \text{Max}(\Gamma_{H_{SM}}|_{m_{\A}},0.075 m_{\A})$ where $\Gamma_{H_{SM}}|_{m_{\A}}$ is the width of a SM Higgs with mass $m_{\A}$ \cite{Heinemeyer:2013tqa}. This accounts for both small Higgs masses for which the width of the peak is caused by detector effects and large Higgs masses for which the physical width dominates.   

\begin{figure}[h!]
\begin{center}
	\includegraphics[scale=0.32]{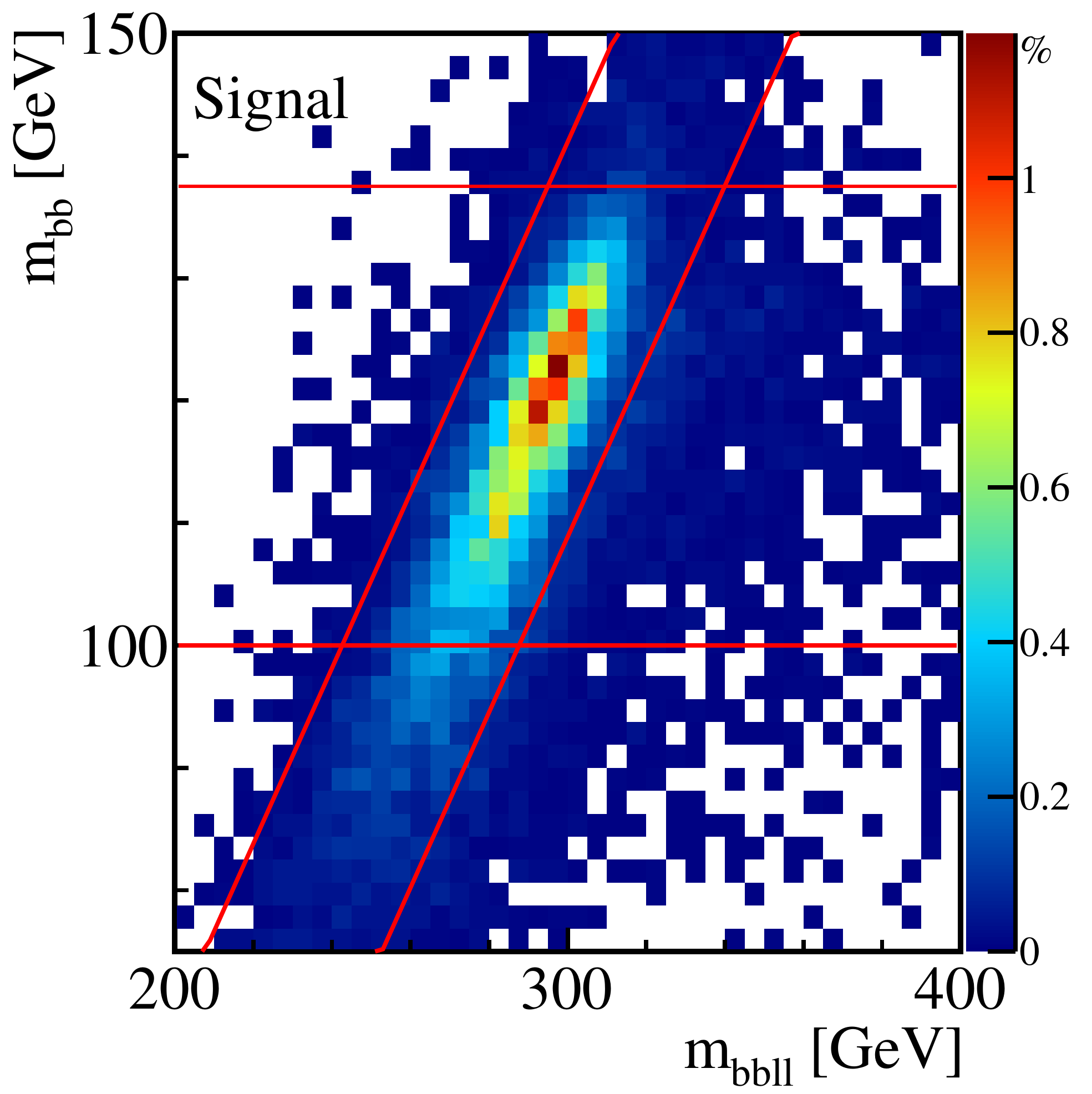}
	\includegraphics[scale=0.32]{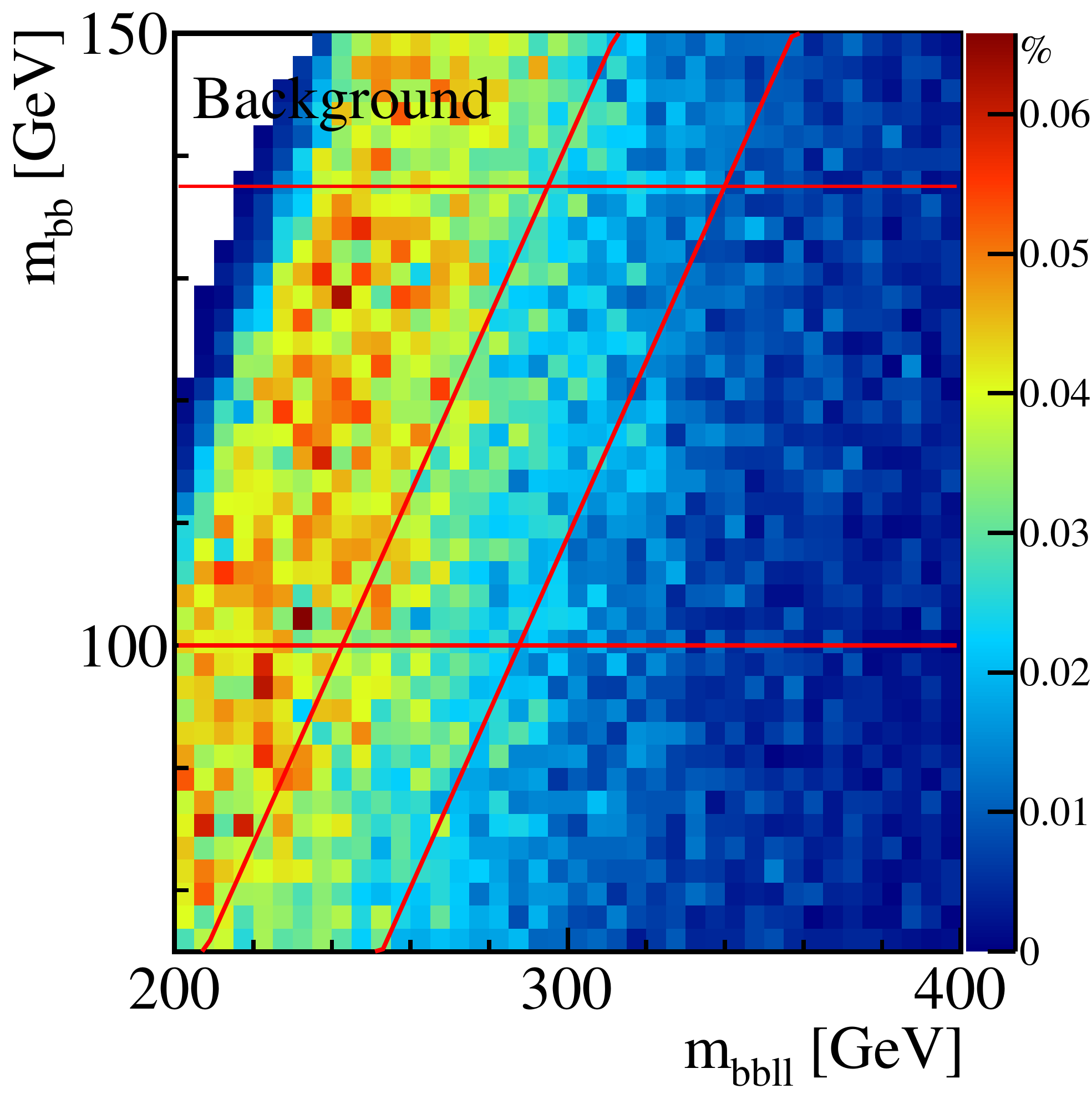}
 \caption{Normalized distribution (in percent as given by the color code along the $y$-axis)  of $m_{bb}$ versus $m_{bb\ell \ell }$ for the signal (left panel), and the backgrounds ($Z/\gamma^{*}bb$+$ZZ$+$H_{SM}$+$t\bar{t}$) (right panel) for $m_{\A}=300$ GeV and $m_{H}=126$ GeV. 
 Two horizontal lines indicate the $m_{bb}$ range and two slanted lines indicate the $m_{bb\ell\ell}$ range,  as given in Eq.~(\ref{cut_4}).   }
\label{fig:2D_mbb_mllbb}
\end{center}
\end{figure}
	
The effectiveness of this cut is shown in Fig.~\ref{fig:2D_mbb_mllbb} for $m_{\A}=300$ GeV and $m_{H}=126$ GeV, with two horizontal lines indicating the $m_{bb}$ range and two slanted lines indicating the $m_{bb\ell\ell}$ range as given in Eq.~(\ref{cut_4}).   Left and right panels show the normalized  distributions for the signal and the backgrounds, respectively.   The color coding is such that points in dark red are most likely, with the probability falling as we reach dark blue as indicated on the right color panel in each plot. The numbers in this panel represent the percentage of the number of events that survive in each bin for the corresponding color.   The signal region in each plot is the region bounded by the two pairs of slanting and horizontal lines. As expected, we see that most of the signal events fall within this strip, while the backgrounds mostly lie outside it. 
 
 	\item \textbf{Transverse momentum:} We require the sum of the transverse momenta of the bottom jets and the sum of the transverse momenta of the bottom jets and leptons to satisfy:
	\begin{equation}
\begin{aligned}
\sum_{b\, \rm{jets}}p_T&>\, 0.6 \times \frac{m_{\A}^2+ m_{H}^2 - m_Z^2}{2 m_{\A}}, \\
\sum_{\ell,\, b\,\rm{jets}}p_T&>\, 0.66 \times  m_{\A}.
\end{aligned}
\label{cut_5}
\end{equation}	
The cuts given in Eq.~(\ref{cut_5}) follow from simple relativistic kinematics applied to the process as applicable to the entire momenta, i.e., $\sum_{b\, \rm{jets}} p_{b_i} =\, \frac{m_{\A}^2+ m_{H}^2 - m_Z^2}{2 m_{\A}}$ assuming that the parent Higgs $\A$ is at rest. We have chosen to specialize this formula to the transverse part alone, including an optimization factor of 0.6.  In Fig.~\ref{fig:pT}, we show how this   $p_T$ cut helps in extracting the signal over the backgrounds for the case where the parent mass is 500 GeV and the daughter mass is 126 GeV. The regions of the plot to the left of the two lines are excluded.   It can be seen that while the signal is   largely intact, a good portion of the backgrounds gets cut out.  
\end{enumerate}
\begin{figure}[h!]
\begin{center}
	\includegraphics[scale=0.32]{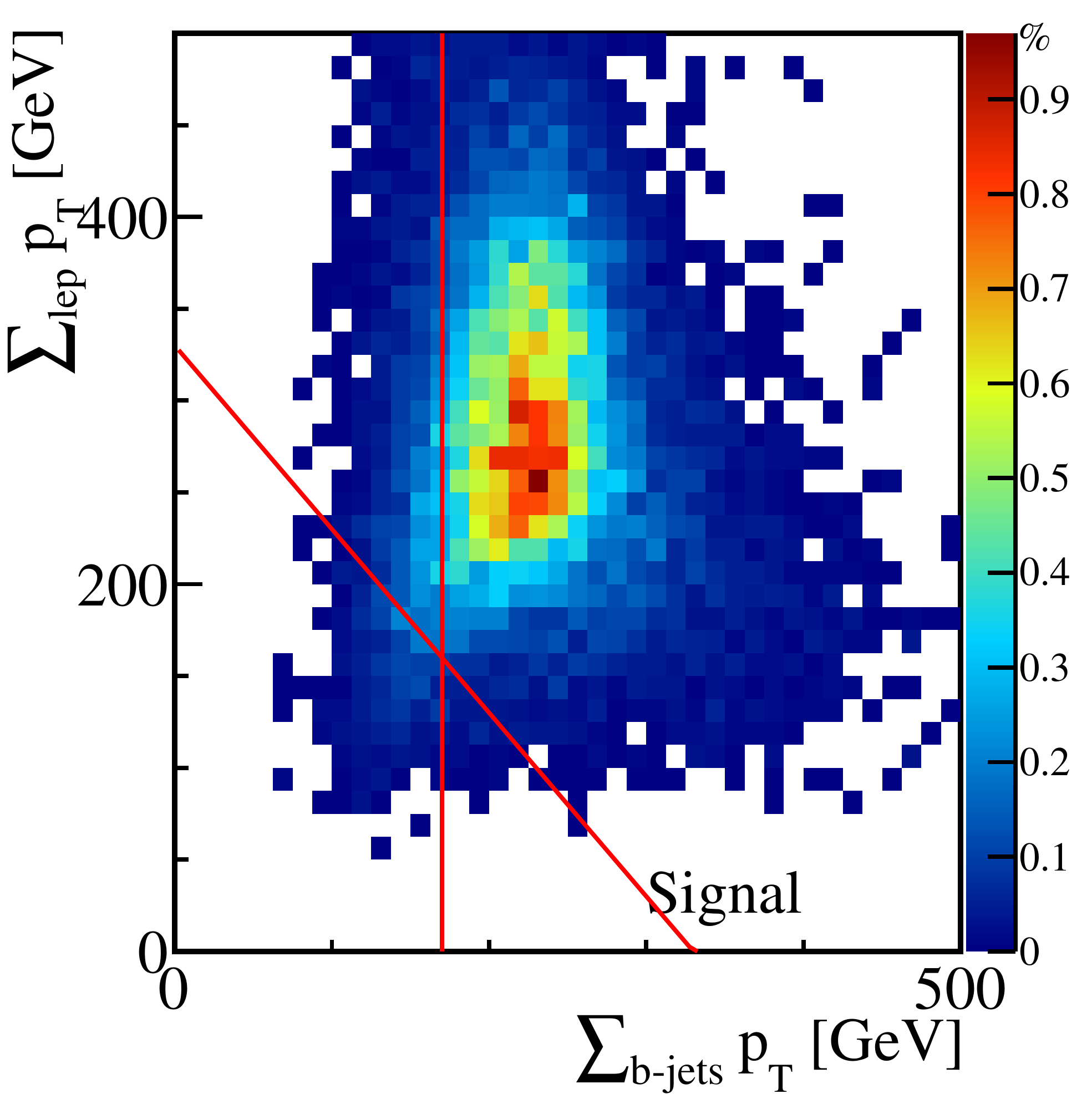}
	\includegraphics[scale=0.32]{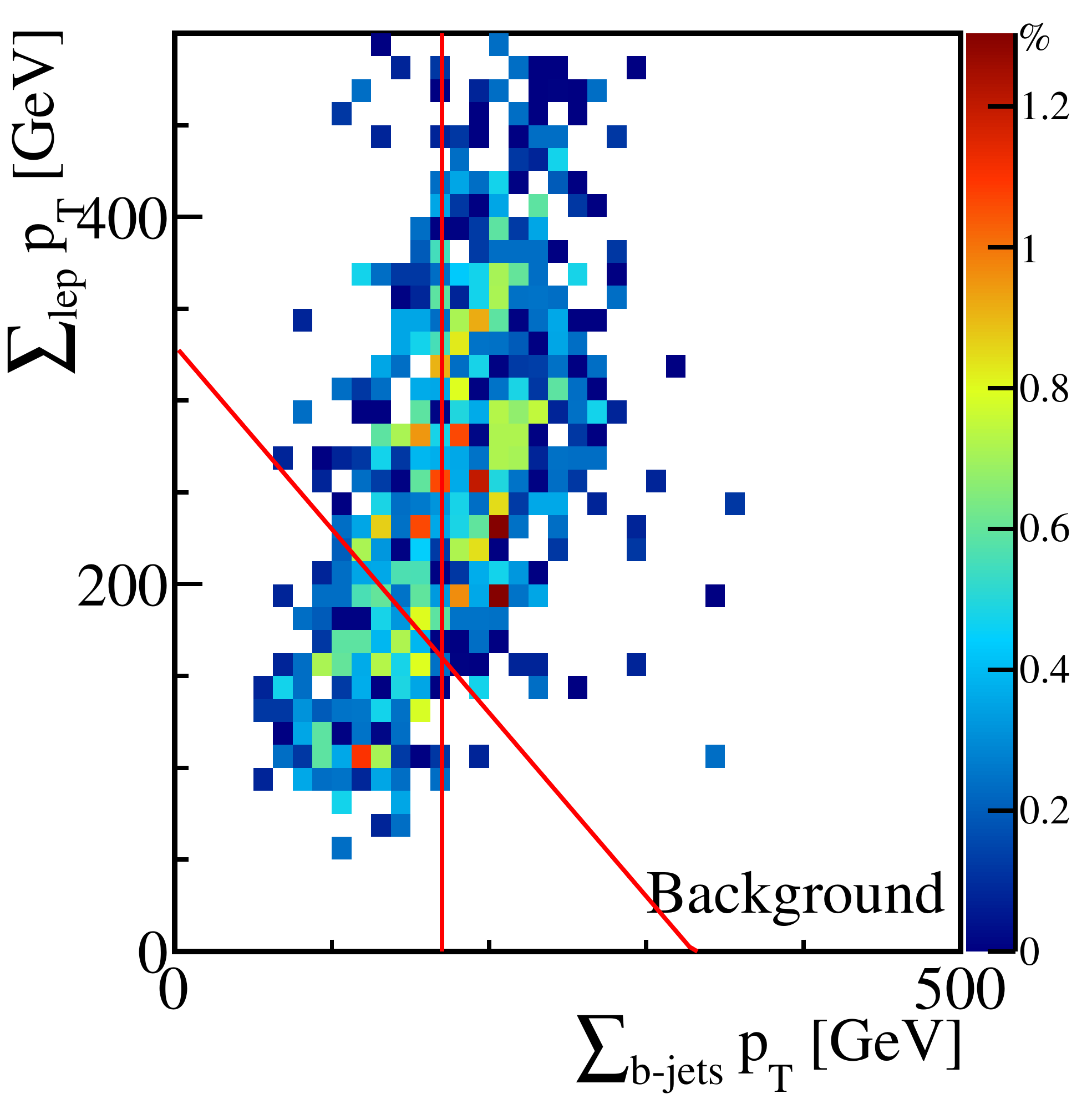}
 \caption{ Normalized transverse momentum distribution  $\sum_{\ell}p_T$ versus $\sum_{b\, \rm{jets}}p_T$  for the signal (left panel) and the backgrounds (right panel) for $m_{\A}$= 500 GeV and $m_{H}$= 126 GeV. Two red lines indicate the conditions used in the cuts as given in Eq.~(\ref{cut_5}).  
 }
\label{fig:pT}
\end{center}
\end{figure}

\begin{table}[h!]
\centering
\resizebox{14cm}{!} {
\begin{tabular}{|l|r|rrrrr|}
\hline
Cut 							&Signal [fb]	&$bb\ell \ell$ [fb]	&$H_{\rm SM}Z$ [fb]	& $t\overline{t}$ [fb] 	&$S/B$		&$S/\sqrt{B}$	\\
\hline
$\sigma_{total}$					&  		&2.21$\times10^6$	&883			&9.20$\times 10^5$	&-		&-		\\
Leptonic decay						&100		&2.21$\times10^6$	&59.4			&2.15$\times 10^4$	&-		&-		\\
Two leptons, Two $b$'s 		[Eq.(\ref{cut_1})]	&6.35 		&343		  	&3.44 			&1409	          	&0.0036 	&2.63		\\
Lepton trigger  		[Eq.(\ref{cut_2})]	&6.35 		&336		  	&3.44 			&1394           	&0.0037 	&2.65		\\
$m_{\ell \ell }$		[Eq.(\ref{cut_3})]	&5.76		&285			&3.13			&189			&0.012 		&4.59		\\
$m_{bb}$ vs $m_{bb\ell \ell}$	[Eq.(\ref{cut_4})]	&3.03 		&11.5			&0.401			&11.5			&0.14 		&11.5		\\
$\sum p_{T,b}$, $\sum (p_{T,b}+p_{T,\ell})$	[Eq.(\ref{cut_5})]	&2.81 	&8.11		&0.361			&8.38			&0.17 		&12.0		\\
\hline
\end{tabular}
}
\caption{Signal and background cross sections with cuts for the signal benchmark point $m_{\A}$ = 300 GeV and $m_{H}$ = 126 GeV at the 14 TeV LHC.  We have chosen a nominal value for $\sigma \times BR(gg \rightarrow \A\to H Z\rightarrow bb\ell\ell)$ of 100 fb to illustrate the cut efficiencies for the signal process.  In the last column,   $S/\sqrt{B}$ is shown for an integrated luminosity of ${\cal L}=300\  {\rm fb}^{-1}$. }
\label{tab:bbcuts}
\end{table}

 In Table \ref{tab:bbcuts}, we show the signal and background cross sections with cuts for signal benchmark point of $m_{\A}$ = 300 GeV and $m_{H}$ = 126 GeV at the 14 TeV LHC.  We have chosen a nominal value for $\sigma \times BR(gg \rightarrow \A/H \rightarrow HZ/ \A Z \rightarrow bb \ell \ell)$ of 100 fb to illustrate the cut efficiencies for the signal process.   In the last column,  $S/\sqrt{B}$ is shown for an integrated luminosity of ${\cal L}=300\  {\rm fb}^{-1}$. Note that for both the signal and the backgrounds, the biggest reduction of the cross sections arises upon demanding exactly two isolated leptons and $b$ jets. In fact, the signal cross section drops from 100 fb to 6.35 fb at this stage. The two $b$ tag efficiencies bring down the cross section by $0.7^2\approx 50\%$. Other contributing factors are leptons and $b$ jets that are either soft or in the forward direction, or non-isolated  leptons and $b$ jets.  We also remark that the $m_{\ell\ell}$ cut does not have a significant effect on   either  the signal or the $bb\ell\ell$ and $H_{\rm SM}Z$ backgrounds since these are dominated by the leptons coming from $Z$, but does have  a pronounced effect on the $t\bar{t}$ background. The second to   last row clearly demonstrates the efficacy of the two dimensional cut in the $m_{bb}-m_{bb\ell\ell}$ plane.   

 \begin{figure}[h!]
	\includegraphics[scale=1.0]{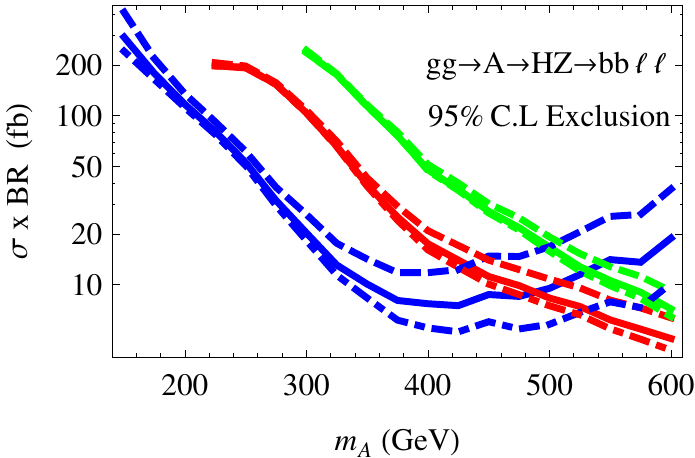}
	\includegraphics[scale=1.02]{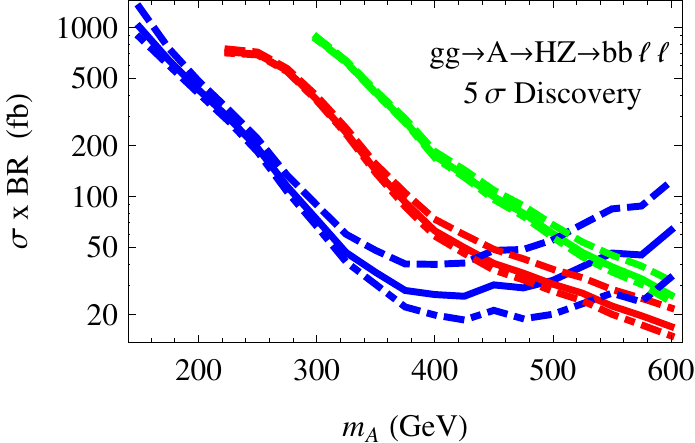}
\caption{The 95\% C.L. exclusion (left) and 5$\sigma$ discovery (right) limits for $\sigma\times{\rm BR}(gg\to\A\to H Z\to bb\ell\ell)$ for $m_{H}=$ 50 GeV (blue), 126 GeV (red), and 200 GeV (green) at the 14 TeV LHC. The dashed, solid and dot-dashed lines correspond to an integrated luminosity of 100, 300 and 1000 fb$^{-1}$,  respectively. Here, we have  assumed a 10\% systematic error on the backgrounds. These results are equally applicable to the $H\to\A Z$ process for the same parent and daughter Higgs masses.   }
\label{fig:bbresults}
\end{figure}

In Fig.~\ref{fig:bbresults}, we display the results at the 14 TeV LHC for 95\% C.L. exclusion (left panel) and 5$\sigma$ discovery (right panel)  limits for $\sigma \times {\rm BR} (gg\to\A\to H Z\to bb\ell\ell)$, which applies for $H\to \A Z$ as well with $m_A$ and $m_H$ switched.   The blue, red, and green curves   correspond to the daughter particle being 50 GeV, 126 GeV, and 200 GeV,  respectively.   The masses of the daughter particle are chosen such that they represent   cases with a light Higgs, a SM-like Higgs, as well as a heavy Higgs  that can decay to $WW/ZZ$.  For each mass, we have displayed the results for three luminosities: 100 fb$^{-1}$ (dashed),  300 fb$^{-1}$ (solid),  and 1000 fb$^{-1}$ (dot-dashed), with 10\% systematic error included \cite{thetaauto}.     Better sensitivity is achieved for larger $m_A$ since the mass cuts on $m_{bb}$ and $m_{bb\ell\ell}$ have  a more pronounced effect on SM backgrounds for larger masses.  The limit, however, gets worse for  the $m_H= 50$ GeV case when  $m_A \gtrsim 400$ GeV (blue curves).  This is due to the decrease of the signal cut efficiency for a highly boosted  daughter particle with two collimated $b$ jets.    For  the interesting case where the daughter particle is 126 GeV, it is seen that the discovery limits for a 300 fb$^{-1}$ collider fall from about 0.7 pb  for $m_A$ of $225$ GeV, to less than 20 fb for a 600 GeV parent particle. These numbers do not change appreciably between the three chosen luminosity values, except for the case of $m_H=50$ GeV and $m_A \gtrsim 400$ GeV.   This is because we have chosen a uniform 10\% systematic error on the backgrounds,  which dominates the statistical errors for most of the parameter region.  For a given parent particle mass $m_A$, limits are better for smaller $m_H=50$ GeV.   This is because the $m_{bb}$ distribution for the dominating $Zbb$ and $tt$ backgrounds peaks around higher masses $m_{bb} \approx $ 70 - 200 GeV and therefore the background rejection efficiency for $m_{bb} \approx$ 50 GeV is high. For $m_{H} = 126$ and $200$ GeV the background rejection efficiencies are comparable but for $m_{H} = 200$ GeV the signal cut efficiency is worse and hence the exclusion  
limits  are the highest for $m_{H} = 200 $ GeV.  

We reiterate here these exclusion and discovery limits are completely model independent. Whether or not discovery/exclusion is actually feasible in this channel should be answered within the context of a particular model, in which the theoretically predicted cross sections and branching fractions can be compared with the exclusion or discovery limits.   We will do this in Sec.~\ref{sec:implications} using Type II 2HDM as a specific  example. 

\subsection{$\A/ H\to  HZ/\A Z\rightarrow \tau\tau\ell\ell$}
\label{sec:tautau}
We now turn to the process $gg\to\A/ H\to HZ /\A Z\rightarrow \tau\tau\ell\ell$. Since we want to reconstruct the final state particles unambiguously, we will employ $\tau$ tags and thus will only consider fully hadronic $\tau$ decays.   While the signal is typically suppressed compared to the $bb \ell \ell$ case due to the smaller $H\rightarrow \tau\tau$ branching fraction, the SM backgrounds \cite{Campbell:2011bn,Dittmaier:2011ti} are much smaller due to the absence of $b$ jets in the final states.  The dominant background is $ZZ$.   We have also included $H_{\textrm{SM}}Z$ background even though it is negligible for most cases.   
 
\begin{figure}[h!]
\begin{centering}
	\includegraphics[scale=0.32]{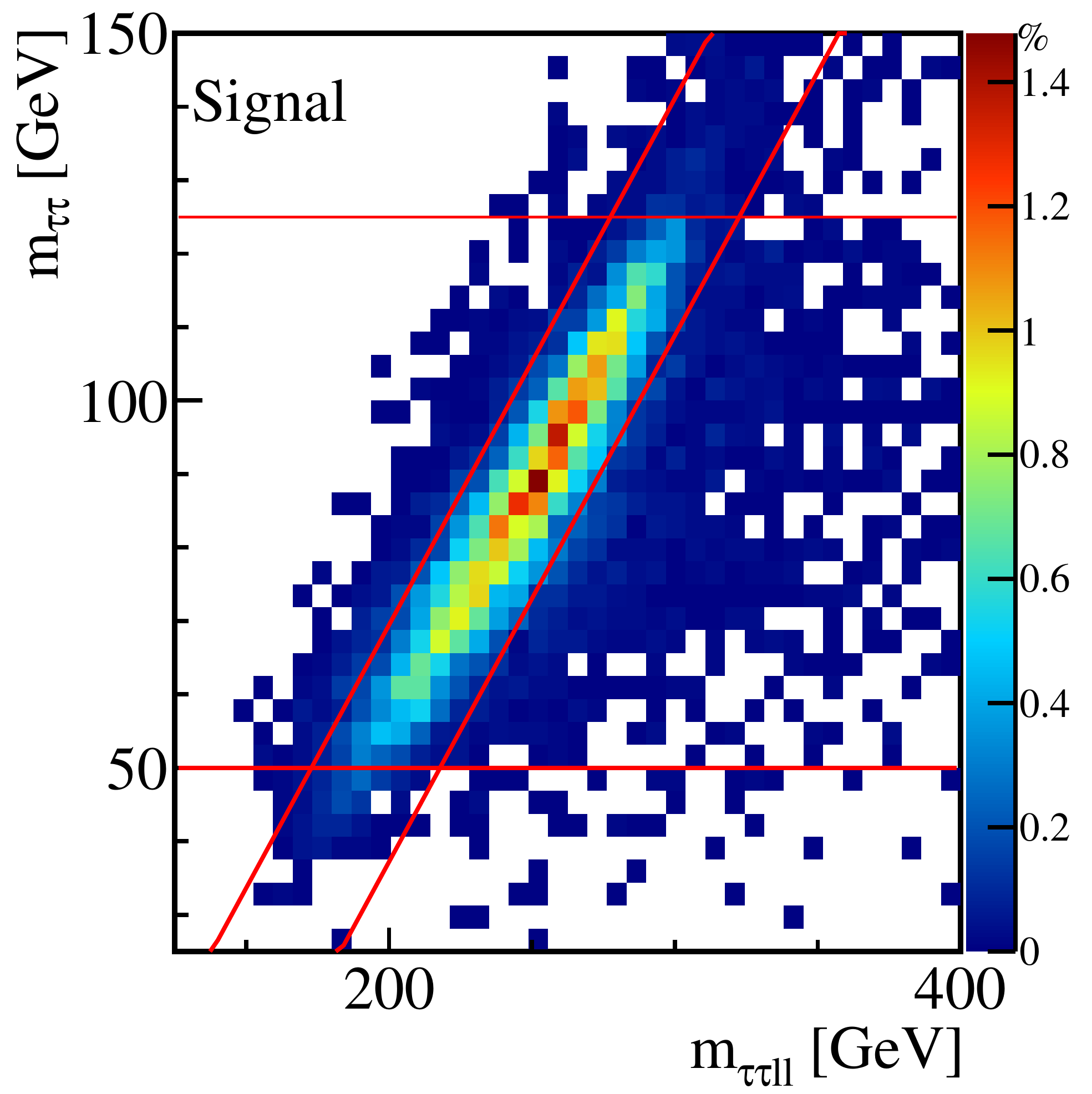}
	\includegraphics[scale=0.32]{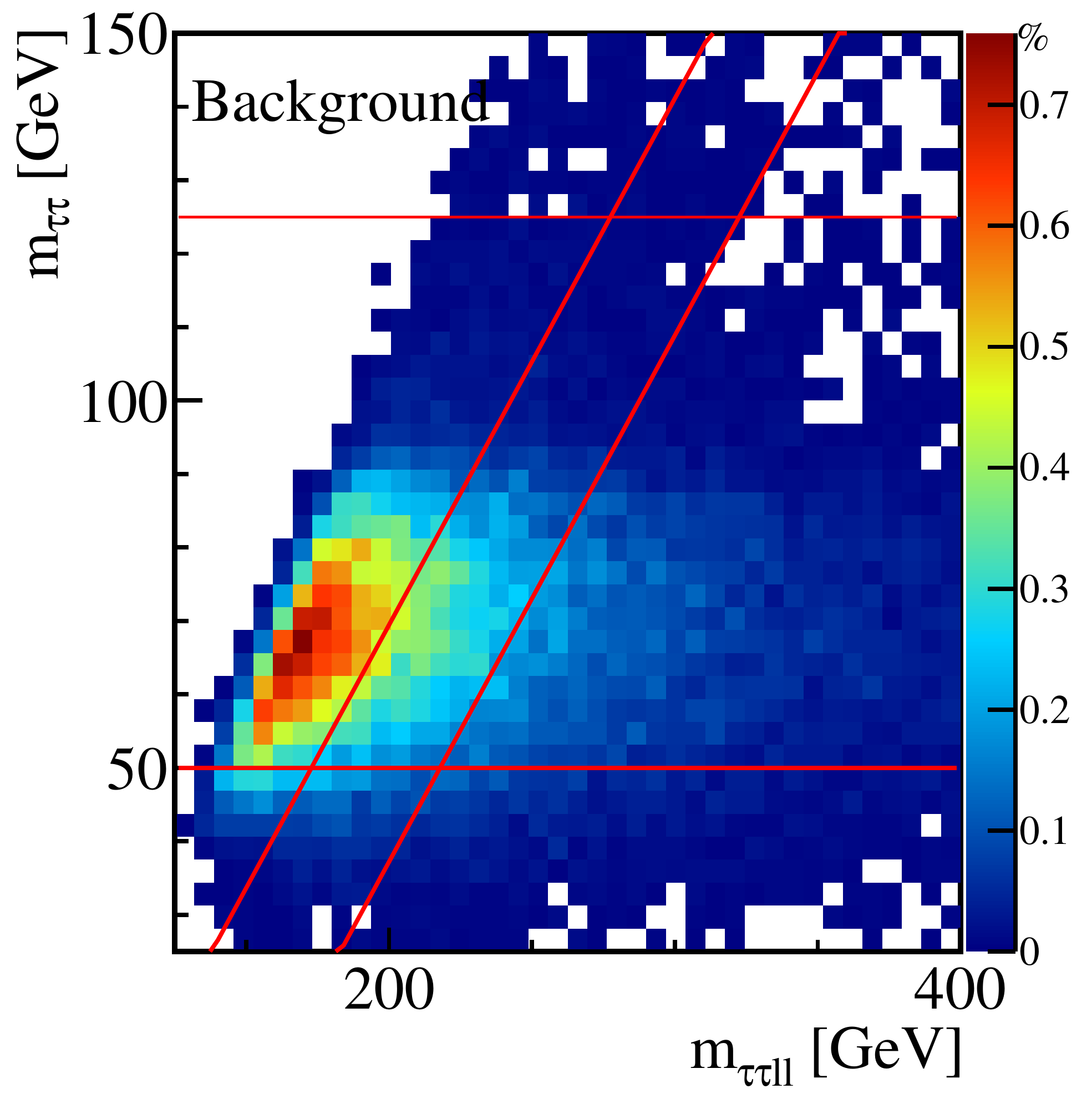}
 \caption{ Normalized distribution of $m_{\tau\tau}$ versus $m_{\tau\tau\ell\ell}$ for the signal (left panel), the  backgrounds (right panel) for $m_{\A}$=300 GeV and $m_H = 126$ GeV. Two horizontal lines indicate the $m_{\tau\tau}$ range and two slanted lines indicate the $m_{\tau\tau\ell\ell}$ range,  as given in Eq.~(\ref{tautau_cut_4}). }
\label{fig6}
\end{centering}
\end{figure}

 Here, we list the cuts employed:
\begin{enumerate}
	\item \textbf{Two isolated leptons and two tagged $\tau$'s:} 
		\begin{equation}
n_{\ell}=2,  \ n_{\tau} =2,\ {\rm with}\ |\eta_{\ell,\tau} |<2.5,\ p_{T,\ell} > 10\ {\rm GeV},\ p_{T, \tau}>20\ {\rm GeV}. 
\label{tautau_cut_1}
\end{equation} 
We do not impose jet veto.
	\item \textbf{Lepton trigger:} 
	\begin{equation}
	p_{T,\ell_1}>30 \text{ GeV}\ \text{or} \ p_{T,\ell_1}>20 \text{ GeV},\ p_{T,\ell_2}>10 \text{ GeV}.
\label{tautau_cut_2}
\end{equation}
	\item \textbf{Dilepton mass $m_{\ell \ell}$:} 
	\begin{equation}
	80 \text{ GeV} <m_{\ell \ell}<  100 \text{ GeV}.
\label{tautau_cut_3}
\end{equation}	
	\item \textbf{$m_{\tau\tau}$ versus $m_{\tau\tau\ell \ell }$:} The expected Higgs mass is shifted more towards smaller values compared to the $bb$ case.  This is because  of  the hadronic decay of  $\tau$ with missing energy carried away by neutrinos. Our 2-D cuts are modified as follows:
 \begin{equation}
\begin{aligned}
(0.7 - w_{\tau\tau})\times m_{H} < \; &m_{\tau\tau} < (0.7 + w_{\tau\tau})\times m_{H} \text{ with } w_{\tau\tau}= 0.3; \\
\frac{m_Z + m_{H}}{m_{\A}}\times(m_{\tau\tau\ell\ell}-  m_{\A} - w_{\tau\tau\ell \ell} )< m_{\tau\tau} & -m_{H} < \frac{m_Z + m_{H}}{m_{\A}}\times(m_{\tau\tau\ell\ell } -m_{\A} + w_{\tau\tau\ell \ell} ), 
\end{aligned}
\label{tautau_cut_4}
\end{equation}		
with $w_{\tau \tau \ell \ell } = \text{Max}(\Gamma_{H_{SM}}|_{m_{\A}},0.075 m_{\A})$. We show the normalized 2-D distribution as well as cuts imposed as indicated by red lines  in Fig.~\ref{fig6} for the signal (left panel) and the  backgrounds (right panel). The cut filters out most of the backgrounds while retaining the signal, yielding a good $S/\sqrt{B}$ value.
	\item \textbf{Transverse momentum:} 
\begin{equation}
\begin{aligned}
\sum_{\tau}p_T>\,  &0.4 \times \frac{m_{\A}^2+ m_{H}^2 - m_Z^2}{2 m_{\A}}, \\
\sum_{\ell,\, \tau }p_T> \, &0.66 \times  m_{\A}.
\end{aligned}
\label{tautau_cut_5}
\end{equation}	
The looser cut on $\sum_{\tau}p_T$ compared to the $bb\ell\ell$ case is again due to the extra missing $E_T$ in the $\tau$ decay.
\end{enumerate}

\begin{table}[htbp]
\centering
\resizebox{14cm}{!} {
\begin{tabular}{|l|r|rrrr|}
\hline
Cut 							&Signal [fb]	&$\tau\tau\ell\ell$ [fb]&$H_{\rm SM}Z$ [fb]	&$S/B$		&$S/\sqrt{B}$	\\
\hline
$\sigma_{total}$					&  		&218			&883		&-		&-		\\
leptonic decay						&10		&218			&3.02		&-		&-		\\
Two leptons, Two $\tau$'s 	[Eq.(\ref{tautau_cut_1})]	&0.43 		&1.622		  	&0.1136		&0.2684	 	&5.921		\\
Lepton trigger  		[Eq.(\ref{tautau_cut_2})]	&0.43 		&1.572		  	&0.1134		&0.2768 	&6.011		\\
$m_{\ell \ell }$		[Eq.(\ref{tautau_cut_3})]	&0.39		&1.312			&0.1031		&0.301	 	&5.869		\\
$m_{\tau\tau}$ vs $m_{\tau\tau\ell \ell }$	[Eq.(\ref{tautau_cut_4})]	&0.29 		&0.3029			&0.023		&0.9643 	&9.192		\\
$\sum p_{T,\tau}$, $\sum (p_{T,\tau}+p_{T,\ell})$[Eq.(\ref{tautau_cut_5})]	&0.18	 	&0.064			&0.013		&2.872 		&12.68		\\
\hline
\end{tabular}
}
\caption{Signal and background cross sections with cuts for signal benchmark point of $m_{\A}$ = 300 GeV and $m_{H}$ = 126 GeV at the 14 TeV LHC.  We have chosen a nominal value for $\sigma \times BR(gg \rightarrow A \rightarrow HZ \rightarrow \tau \tau \ell \ell)$ of 10 fb to illustrate the cut efficiencies for the signal process.  In the last column,  $S/\sqrt{B}$ is shown for an integrated luminosity of ${\cal L}=300\  {\rm fb}^{-1}$.  }
\label{tab:tautaucuts}
\end{table}

In Table \ref{tab:tautaucuts}, we present the cross sections after the individual cut is imposed sequentially. We take a nominal signal cross section of 10 fb to illustrate the efficiency of the chosen cuts.  Again,   the 2-D $m_{\tau\tau}-m_{\tau\tau\ell\ell}$ cut improves the $S/\sqrt{B}$ value significantly.

\begin{figure}[h!]
	\includegraphics[scale=1.0]{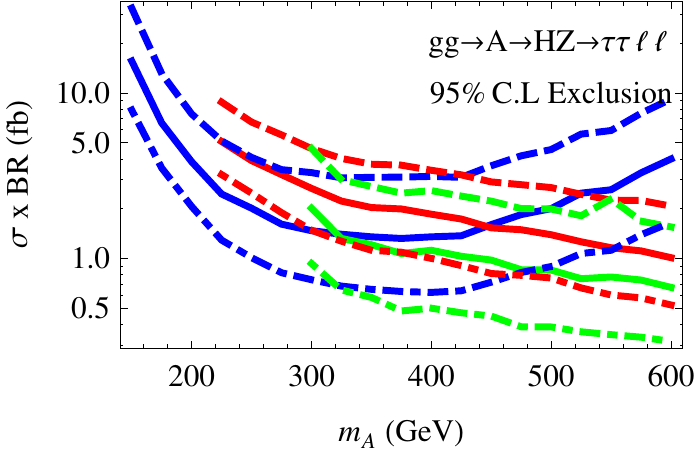}
	\includegraphics[scale=1.0]{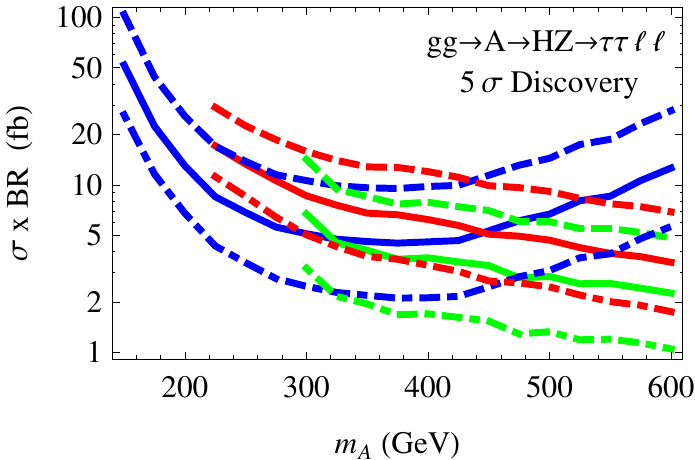}
\caption{The 95\% C.L. exclusion (left) and 5$\sigma$ discovery (right) limits  for $\sigma\times{\rm BR}(gg\to\A\to H Z\to\tau\tau\ell\ell)$ for $m_{H}=$ 50 GeV (blue), 126 GeV (red), and 200 GeV (green) at  the 14 TeV LHC. The dashed, solid and dot-dashed lines correspond to an integrated luminosity of 100, 300 and 1000 fb$^{-1}$,  respectively.    A 10\% systematic error on the backgrounds is assumed as well.  }
\label{fig:ttresults}
\end{figure}

In Fig.~\ref{fig:ttresults}, we show the 95\% C.L. exclusion and 5$\sigma$ discovery reach in $\sigma\times{\rm BR}(gg\rightarrow \A \rightarrow H Z \rightarrow \tau\tau\ell\ell)$ for the 14 TeV LHC. The general feature of these plots follows that of Fig.~\ref{fig:bbresults}, particularly with highly boosted daughter particles making $\tau$ identification more challenging, as shown by the blue curves for 50 GeV daughter particle mass, which exhibit worse limits for $m_A > 400$ GeV.   The exclusion limits are lowest for small $m_H = 50$ GeV and also for high $m_H = 200$ GeV since the dominating $ZZ$ background peaks at $m_{\tau\tau} \approx 90$ GeV and therefore our $m_{\tau \tau}$ mass cut leads to a high background rejection for lower or higher $m_H$.    Since the statistical error dominates   the 10\% systematic error, the $\sigma\times{\rm BR}$ limits scale roughly with $1/\sqrt{L}$, as indicated by the dashed, solid and dot-dashed lines for different luminosities.

Compared to the $bb\ell\ell$ case, the $\sigma\times{\rm BR}$ reach in $\tau\tau\ell\ell$ case is better due to significantly lower SM backgrounds.   For the 126 GeV daughter particle case with  300 fb$^{-1}$, the 5$\sigma$ discovery reach varies from about 20 fb for parent mass of 225 GeV  to about 3 fb for 600 GeV. Thus,   given the typical ratio of ${\rm Br}(H/\A \rightarrow bb ):{\rm Br}(H/\A \rightarrow \tau\tau) \sim 3 m_{b}^2/m_\tau^2  $, the reach in $\tau\tau\ell\ell$ can be comparable or even better than $bb\ell\ell$ channel, in particular,  for smaller parent Higgs masses.

\subsection{$\A\to H Z\rightarrow ZZZ\rightarrow 4\ell+2j$}
\label{sec:ZZZ}
We now consider the case where the daughter particle decays to a pair of $Z$ bosons, which only applies to $\A\to H Z\rightarrow ZZZ$. This process involves a trade-off between having a clean final state with suppressed backgrounds and suppressed signal cross section for detection.  We find that the best final states combination that yields signal cross sections that are not too suppressed in realistic models with controllable backgrounds is the $4\ell+2j$ final state: $\A \rightarrow HZ \rightarrow ZZZ \rightarrow 4\ell+2j$.   The SM backgrounds for this process come from the single, double and triple vector boson processes including additional jets as well as $t\bar{t}$ background  \cite{snowmassdetector, Avetisyan:2013onh, Avetisyan:2013dta}.
 
Note that the $Z$'s from the $H$ decay could be either on-shell or off-shell depending on $m_{H}$.  We will display our results for two cases: one where one of the final state $Z$'s is necessarily off-shell, and another where both are on-shell. We will find that the latter case leads to much better discovery prospects. 

We applied the following set of cuts:
\begin{itemize}
	\item \textbf{Four isolated leptons, two jets:}  
	\begin{equation}
n_{\ell}=4, \ n_{j} \geq 2,\ {\rm with}\ |\eta_{\ell} |<2.5,\ p_{T,\ell} > 10\ {\rm GeV},|\eta_{j} |<5,\ p_{T,j} > 20\ {\rm GeV}. 
\label{eq:ZZ_cut_1}
\end{equation}
 	For jet reconstruction, we use the anti-$k_T$ jet algorithm with $R= 0.5$.    We also require the leptons to satisfy the lepton trigger as in Eq.~(\ref{cut_2}).

	\item \textbf{Three $Z$-candidates:} We reconstruct the hadronically decaying $Z$ using the 2 hardest jets. To reconstruct the leptonically decaying $Z$'s: 
	\begin{itemize}
	 \item \textbf{$4e$ or $4\mu$:} If we have 4$e$ or 4$\mu$,  we first find the combination of electrons or muons  with opposite charge that is closest to the $Z$-mass. The other 2 electrons or muons are combined to find the last $Z$. 
 	 \item \textbf{2$e$2$\mu$:} Here, we combine the same flavored leptons in a straightforward manner.	 
	\end{itemize}	
	
	\item \textbf{$Z$ masses:} We require the hardonically decaying $Z_1$, the well reconstructed leptonically decaying $Z_2$ and the final reconstructed leptonically decaying $Z_3$ to be in the following windows:
\begin{equation}
\begin{aligned}
60 \text{ GeV} <& m_{Z_1}<  115 \text{ GeV}.\\
80 \text{ GeV} <& m_{Z_2}<  100 \text{ GeV}.\\
m_{\text{min}} <& m_{Z_3}<  115 \text{ GeV}.
\end{aligned}
\label{eq:ZZ_cut_3}
\end{equation}
Here, we assume $Z_1$ to be on-shell. However, we allow for the possibility that $Z_3$ could be far off-shell. The $m_{\textrm{min}}$ employed here mimics the LHC search strategy for the SM Higgs, and its value depends on the Higgs mass and can be found in Table 2 of Ref.~\cite{ATLAS:2012ac}. 
	
	\item \textbf{$m_{H}$ and $m_{\A}$:} The $Z$ produced in the $\A$ decay typically has   a higher $p_T$ than the $Z$'s produced in $H$ decay. Therefore we assume that   the lower $p_T$ $Z$'s are coming from the $H$. For the reconstructed $H$ with mass $m_{ZZ}$ and $\A$ with mass $m_{ZZZ}$ we require:  
\begin{eqnarray}
 0.9 \ m_{H} <& m_{ZZ}&<  1.1 \ m_{H} \label{eq:ZZ_cut_4_1}
\\ 
0.875 \ m_{\A} <& m_{ZZZ}&<  1.125 \ m_{\A}.
 \label{eq:ZZ_cut_4_2}
\end{eqnarray}	
\end{itemize}

In Table \ref{tab:ZZcuts}, we show the cross sections after cuts for two signal benchmark points $m_H=$ 126 GeV and 200 GeV with $m_A$ fixed at  400 GeV,   as well as for the SM backgrounds.   For   $m_{H}$ = 126 GeV,  we choose a signal cross section of 1 fb.\footnote{Particularly, the number is arrived at by taking gluon fusion cross section of 9 pb for a 400 GeV CP-odd Higgs, and assuming BR($\A\to H Z$)= 50\% and ${\rm Br}(H\rightarrow ZZ^*)$=2.64\% for a 126 GeV Higgs.   }    For $m_{H}=$ 200 GeV, we use a cross section of 10 fb assuming BR$(H\to ZZ)$ is 25\% for $m_{H}=$ 200 GeV.  For the  $m_H=$126 GeV  case, due to the off-shell $Z$ decay, the cut efficiencies for identifying four leptons and two jets, as well as reconstructed $m_{ZZ}$ cuts are fairly low.  Coupled with the small branching fraction of $H \rightarrow ZZ^*$,  the number of surviving events is about  1 for 100 fb$^{-1}$ after all cuts are imposed.    However, this channel becomes quite promising for heavier  daughter masses when all $Z$'s in the final state are on-shell, as shown for the benchmark point of $m_H=$ 200 GeV.

   \begin{table}[htbp]
\centering
\resizebox{14cm}{!} {
\begin{tabular}{|l|r r|rrr|}
\hline
Cut                             &  $m_H=$126 GeV & $m_H=$ 200 GeV   &BG [fb]        &$S/B$        &$S/\sqrt{B}$    \\
\hline
Leptonic decay                       &1.0 &10        &            &        &-        \\
Four leptons, Two jets     [Eq.(\ref{eq:ZZ_cut_1})]   &0.14 &2.78         &           2.592     &     1.07     &      29.9    \\
$Z$-mass              [Eq.(\ref{eq:ZZ_cut_3})]    &0.027&1.03     &          0.6027     &     1.71     &      23.1    \\
$m_{ZZ}$            [Eq.(\ref{eq:ZZ_cut_4_1})]    &0.012&0.73     &          0.2118     &      3.49    &      27.9    \\
$m_{ZZZ}$            [Eq.(\ref{eq:ZZ_cut_4_2})]    &0.0094&0.54     &         0.0905     &      5.98     &      31.2    \\
\hline
\end{tabular}
}
\caption{Signal and background cross sections with cuts for signal benchmark point of $m_{\A}$ = 400 GeV and $m_{H}$ = 126 or 200 GeV at the 14 TeV LHC.  We have chosen a nominal value for $\sigma \times BR(gg \rightarrow \A \rightarrow H Z \rightarrow ZZZ\rightarrow 4\ell+2j)$ of 1.0 fb (for 126 GeV $m_H$) and 10 fb (for 200 GeV $m_H$)  to illustrate the cut efficiencies for the signal process.  The total background cross section after cuts is shown by imposing the cuts for the $m_H = 200$ GeV case. $S/B$ and $S/\sqrt{B}$ are given for the $m_H=200$ GeV benchmark point.  In the last column,   $S/\sqrt{B}$ is shown for an integrated luminosity of ${\cal L}=300\  {\rm fb}^{-1}$.}
\label{tab:ZZcuts}
\end{table}

We note that the nominal value for the  cross section that is used in  Table \ref{tab:ZZcuts} can, in typical BSM scenarios, be enhanced at small   $\tan\beta$, due to the top loop  contributions to the gluon fusion production, as well as the suppression of the $H \rightarrow bb$ branching fraction.    

\begin{figure}[h!]
\begin{centering}
	\includegraphics[scale=1.0]{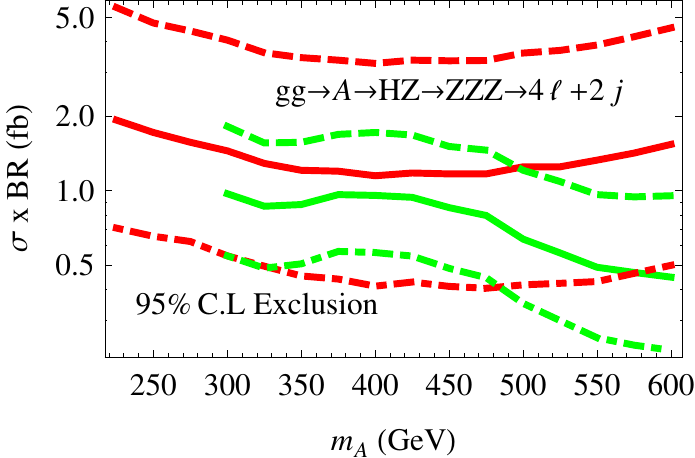}
	\includegraphics[scale=1.02]{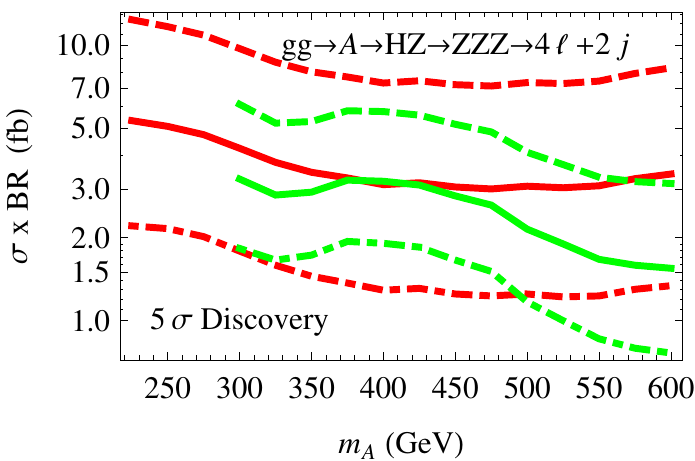}
\caption{The 95\% C.L. discovery and 5$\sigma$ exclusion limits at the 14 TeV LHC in the channel $gg\to\A\to H Z\to ZZZ\to 4\ell+2j$ for $m_{H}=$ 126 GeV (red) and $m_{H}=$ 200 GeV (green).   The dashed, solid and dot-dashed lines correspond to an integrated luminosity of 100, 300 and 1000 fb$^{-1}$,  respectively.    A 10\% systematic error on the backgrounds is assumed as well.     
 }
\label{fig:zzresults}
\end{centering}
\end{figure}

Fig.~\ref{fig:zzresults} shows the 95\% C.L.  exclusion and 5$\sigma$ discovery at the 14 TeV LHC for different integrated luminosities: $\mathcal{L}=$ 100 fb$^{-1}$, 300 fb$^{-1}$, and 1000 fb$^{-1}$.  Even for $\mathcal{L}=$ 300 fb$^{-1}$, the discovery limits vary only between about 3 fb and 1.5 fb with 200 GeV $m_H$ for $m_{\A}$ between 300 GeV and 600 GeV. Thus, the only challenge in this channel is to have high enough signal cross sections, as the SM backgrounds prove to be less of a threat compared to the $bb\ell\ell$ final state.


\section{Implications for the Type II 2HDM}
\label{sec:implications}

The decays $A/H \rightarrow HZ /AZ$ appear in many models that have an extension of the SM Higgs sector.  In this section, we illustrate the implications of the exclusion or discovery limits of $bb\ell\ell$, $\tau\tau\ell\ell$ and $ZZZ (4\ell 2j)$ searches on these models using Type II 2HDM as an explicit example.

In the Type II 2HDM,  one Higgs doublet $\Phi_1$ provides masses for  the down-type quarks and charged leptons, while the other Higgs doublet $\Phi_2$ provides masses for the up-type quarks.   The couplings of the  CP-even Higgses $\h$, $\H$ and the CP-odd Higgs $\A$ to the SM gauge bosons and fermions are scaled by a factor $\xi$ relative to the SM   value, which are presented in Table~\ref{tab:couplings}.

\begin{table}[h]
\begin{center}
  \begin{tabular}{| l | p{2.5cm} || l | p{2.5cm} || l | p{1.5cm}| }
    \hline
    $\xi^{VV}_{\h}$ & $\sin(\beta-\alpha)$ & $\xi^{VV}_{\H}$ &$\cos(\beta-\alpha)$& $\xi^{VV}_{\A}$ &$0$ \\ \hline
    $\xi^{u}_{\h}$ & $\cos\alpha/\sin\beta$ & $\xi^{u}_{\H}$ & $\sin\alpha/\sin\beta$& $\xi^{u}_{\A}$ & $\cot\beta$\\ \hline
    $\xi^{d,l}_{\h}$ & $-\sin\alpha/\cos\beta$ & $\xi^{d,l}_{\H}$ & $\cos\alpha/\cos\beta$& $\xi^{d,l}_{\A}$ & $\tan\beta$\\ \hline
  \end{tabular}
\end{center}
\caption{The multiplicative factors $\xi$ by which the couplings of the CP-even Higgses and the CP-odd Higgs  to the gauge bosons and fermions scale with respect to the SM value.  The superscripts $u,d,l$ and $VV$ refer to the up-type quarks, down-type quarks, leptons, and $WW/ZZ$ respectively. }
\label{tab:couplings}
\end{table}

The implication of the current Higgs search results on the Type II 2HDM has been studied in the literature \cite{Craig:2012vn, Chiang:2013ixa, Grinstein:2013npa, Coleppa:2013dya,  2HDM_other}.  In particular,  a detailed analysis of the surviving regions of the Type II 2HDM  was performed in \cite{Coleppa:2013dya}, considering various theoretical constraints and including  the latest experimental results from both the ATLAS and the CMS.  Either the light or the heavy CP-even Higgs can be interpreted as the observed 126 GeV SM-like Higgs, with very different preferred parameter regions.   In the $\h$-126 case, we are restricted   to narrow regions with $\sba\sim\pm$ 1 with $\tan\beta$ up to 4 or an extended region in $0.55 < \sba < 0.9$ with $1.5 < \tan\beta<4$. The masses $m_{\H}, m_{H^\pm}$, and $m_{\A}$ are, however, relatively unconstrained. In the $\H$-126 case, we are restricted to a narrow region of $\sba\sim$ 0 with $\tan\beta$ up to about 8, or an extended region of $\sba$ between $-0.8$ to $-0.05$,   with $\tan\beta$ extending to 30 or higher. $m_{\A}$  and $m_{H^\pm}$ are nearly degenerate due to $\Delta\rho$ constraints.  Imposing the flavor constraints in addition   further narrows down the preferred parameter space.

Given the different parameter dependence of the gluon fusion cross section for $\A$ and $\H$, the branching fractions of $\h$, $\H$ and $A$,  as well as the coupling difference between $\h\A Z$ and $\H\A Z$,  we can identify three different classes of processes: $gg\rightarrow \A \rightarrow \h Z$,  $gg\rightarrow \A \rightarrow \H Z$, and $gg\rightarrow \H \rightarrow \A Z$ when interpreting the exclusion and discovery limits from the previous sections.  We do not consider the decay of $\h \rightarrow \A Z$ since this channel is experimentally challenging given that both $\h$ and $\A$ are relatively light.

\begin{table}[h]
\begin{center}
  \begin{tabular}{|l|c|c|c|c| }
    \hline
    $\left\{ {m_{\A},m_{\H},m_{\h}}\right\}$ GeV & $\A\to\h Z$ & $\A\to\H Z$ &$\H\to\A Z$& Favored Region  \\ \hline
   BP1: $\left\{ {400, 126, 50}\right\}$ & \cmark & \cmark & \xmark& $\sba\approx$ 0 \\ \hline
    BP2: $\left\{ {400, 200, 126}\right\}$ & \cmark & \cmark & \xmark& $\sba\approx\pm$ 1 \\ \hline
     BP3: $\left\{ {300, 400, 126}\right\}$ & \cmark & \xmark & Marginal& $\sba\approx\pm$ 1 \\ \hline
    BP4: $\left\{ {50, 400, 126}\right\}$ & \xmark & \xmark & \cmark& $\sba\approx\pm$ 1 \\ \hline
    BP5: $\left\{ {200, 400, 126}\right\}$ & \xmark & \xmark & \cmark& $\sba\approx\pm$ 1 \\ \hline
   \end{tabular}
\end{center}
\caption{Benchmark points shown for illustrating the discovery and exclusion limits in the processes considered in the context of the Type II 2HDM. The checkmarks indicate kinematically allowed channels. Also shown are the typical favored region of $\sin(\beta-\alpha)$  for each case (see Ref.~\cite{Coleppa:2013dya}).  }  
\label{tab:classification}
\end{table}
 
In Table.~\ref{tab:classification}, we list the benchmark points  that we use for the interpretation of the exclusion and discovery bounds in the Type II 2HDM.    BP1 is the only $\H$-126 case   while BP2$-$BP5 are for the $\h$-126 case.   Both BP1 with $(m_{\A},m_{\H},m_{\h}) =  (400, 126, 50)$ GeV and BP2 with $(m_{\A},m_{\H},m_{\h}) =  (400, 200, 126)$ GeV are  designed for both $gg \rightarrow \A \rightarrow \H Z$ and  $gg \rightarrow  \A \rightarrow \h Z$  as both modes are kinematically open.    BP2 with $m_{\H}=200$ GeV, in particular,   allow us to study the implication of    $ZZZ (4 \ell 2j)$ search through $gg \rightarrow  \A \rightarrow \H Z$.   BP3 with $(m_{\A},m_{\H},m_{\h})=  (300, 400, 126)$ GeV is designed for $\A \rightarrow \h Z$ with the $\H$ decoupled.   We also choose $m_A$ to be below the $t\bar{t}$ threshold.    BP4 with $(m_{\A},m_{\H},m_{\h})=  (50, 400, 126)$ GeV  and BP5 with $(m_{\A},m_{\H},m_{\h})=  (200, 400, 126)$ GeV are  designed for the study of $gg \rightarrow \H \rightarrow \A Z$.  Also shown in Table~\ref{tab:classification} are  the preferred regions in $\sin(\beta-\alpha)$   once all the theoretical and experimental constraints are imposed, following Ref.~\cite{Coleppa:2013dya}.    

Note that   in our study, we have decoupled the charged Higgs so that it does not appear in the decay products of $A$ or $H$.  For a light charged Higgs that is accessible in the decays of $A/H \rightarrow H^\pm W^\mp, H^+H^-$, decay branching fractions of $A/H \rightarrow HZ /AZ$ will decrease correspondingly, which reduces the reach of this channel.   However, the new decay channels involving the charged Higgs might provide new discovery modes for $A$ or $H$, which have been explored elsewhere \cite{Tong_Su,other_Hpm,Coleppa_Kling_Su}.   In particular, for $A/H \rightarrow H^\pm W^\mp, H^+H^-$ with $H^\pm \rightarrow \tau^\pm \nu$,  the spin correlation in the  $\tau$ decay can be used to identify the signal from the SM backgrounds.   The sensitivity of this channel involving $H^\pm$ in the intermediate to large $\tan\beta$ region provides a nice complementarity to the $A/H \rightarrow HZ /AZ$ channels~\cite{Tong_Su}.

To be more general, in the discussion below when we interpret the search results of $bb\ell\ell$, $\tau\tau\ell\ell$ and $ZZZ (4\ell 2j)$ channels in the model parameter space,  we do not restrict ourselves to  the narrow preferred parameter regions for $\h$-126 or $\H$-126 case as shown in Ref.~\cite{Coleppa:2013dya}.   In particular, we consider the broad range of  $-1\leq\sba\leq+1$ and $1\leq \tan\beta\leq 50$.  This is because  the allowed regions would change if a soft $\mathcal{Z}_2$ symmetry breaking  is incorporated which Ref.~\cite{Coleppa:2013dya} did not deal with.  Furthermore,  the Higgs sector of 2HDM and the subsequent symmetry breaking structure is rather general and the results presented in this section can   be interpreted in the context of any such model if   the Higgs couplings to the fermions follow a similar pattern.  We do, however, point out the interplay between the exotic Higgs decay channels  and the SM-like Higgs search results at the end of each discussion.     
 
\subsection{$gg\to\A\to\h Z$}

\begin{figure}[h!]
\centering
	\includegraphics[width=2.5 in]{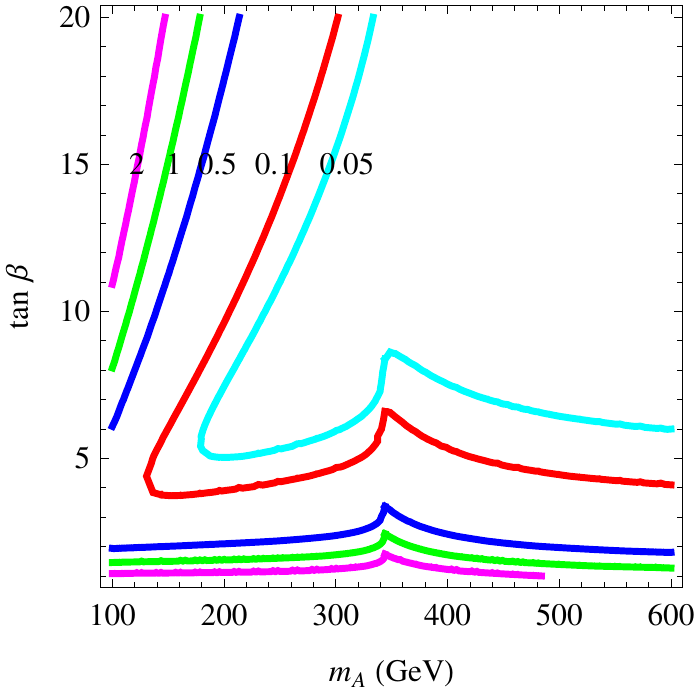}
	\includegraphics[width=2.5 in]{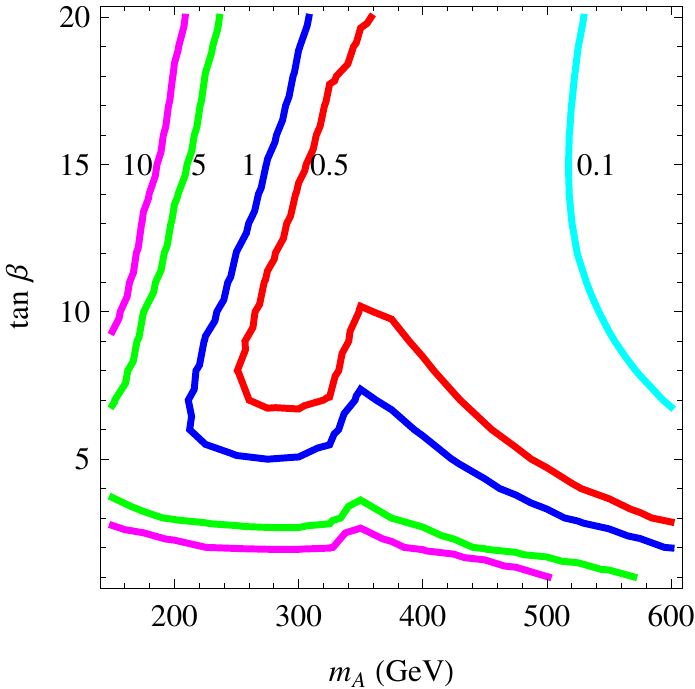}
\caption{Contours of $\sigma(gg\to\A)$ normalized to the SM value  in the $m_{\A}-\tan\beta$ plane (left panel) and  $\sigma(gg\to\A)$ at the 14 TeV LHC in unit of pb (right panel).  }
\label{fig:SigmaAfull}
\end{figure}

We compute the production cross section for the CP-odd Higgs $\A$ by a simple rescaling of the SM Higgs cross section as follows:
\begin{equation}
 \sigma(gg\rightarrow \A)={\sigma_{\rm SM}}\times \frac{|\cot\beta\, F^{A}_{1/2}(\tau_t)+\tan\beta\, F^{A}_{1/2}(\tau_b)|^2}{|F^{h}_{1/2}(\tau_t)+F^{h}_{1/2}(\tau_b)|^2},
 \label{eq:sigmaA}
\end{equation}
where $\tau_f=4m_f^2/m_A^2$ and the scalar and pseudoscalar loop factors  $F^{h}_{1/2}$ and $F^{A}_{1/2}$ are given by:  \cite{Gunion:1989we}
\begin{equation}
 F^{A}_{1/2}=-2\tau f(\tau),\ \ \ 
 F^{h}_{1/2}=-2\tau\left[1+(1-\tau)f(\tau) \right],
 \label{eq:loopfactors}
\end{equation}
and 
\begin{equation}
	f(\tau) = \left\{ \begin{array}{lc} 
	\left[ \sin^{-1} (1/\sqrt{\tau}) \right]^2 \ \ & \tau \geq 1, \\   
	-\frac{1}{4} \left[ \ln(\eta_+/\eta_-) - i \pi \right]^2 \ \ & \tau < 1,
	\end{array} \right.
\end{equation}
with $\eta_{\pm} \equiv 1 \pm \sqrt{1 - \tau}$.  We have ignored the contribution from other Higgses in the loop, which is typically small.  The left panel of Fig.~\ref{fig:SigmaAfull} shows the contour plot of the   $\sigma(gg \rightarrow \A)$ normalized to that of the  SM Higgs with the same mass.  The $\tan\beta$ dependence is due to  the $\A tt$ and $\A bb$ couplings, while the mass dependence   comes from the different dependence of $F_{1/2}(\tau_f)$  on $\tau_f$ for pseudoscalar compared to a scalar.   Enhancements over the SM value is possible for  large $\tan\beta$ at small $m_A$ due to the bottom loop, or small $\tan\beta$ for all values of $m_A$ due to the top loop.    The bump in the plot for $m_A$  around 350 GeV corresponds to  top threshold effects. Note   that for $\A$, the production cross section only depends on $\tan\beta$ and is  independent of   $\alpha$. Also shown in  the right panel of Fig.~\ref{fig:SigmaAfull} are contours of $\sigma(gg\to\A)$ in the $m_{\A}-\tan\beta$ plane  for the 14 TeV LHC, with the cross sections for the SM Higgs production obtained from Ref.~\cite{Heinemeyer:2013tqa,Hahn:2010te}.   Significant cross sections of 10 pb or more are possible for  large $m_A$ up to 500 GeV for small $\tan\beta$. Cross sections of similar magnitude are also possible at large $\tan\beta$ due to the bottom loop enhancement effects, albeit  only for relatively small $m_A$.     

\begin{figure}[h!]
\centering
	\includegraphics[width=2.5in]{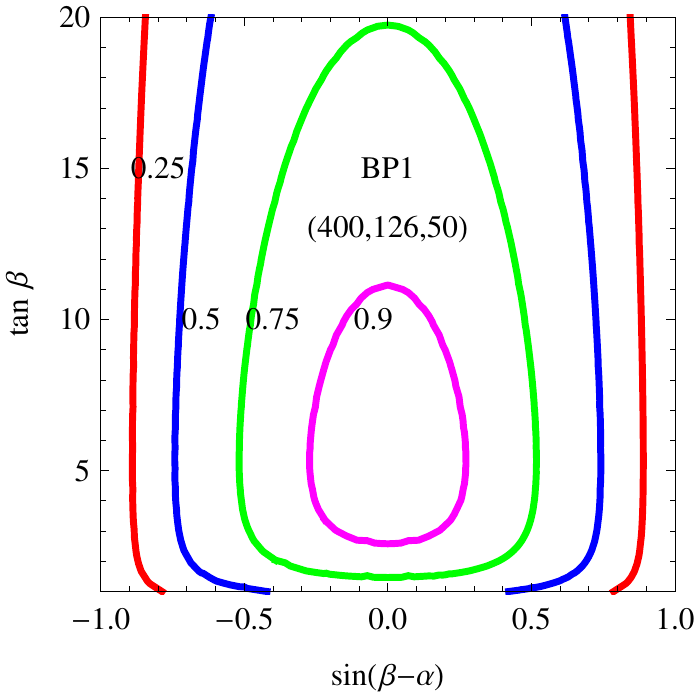}
	\includegraphics[width=2.5in]{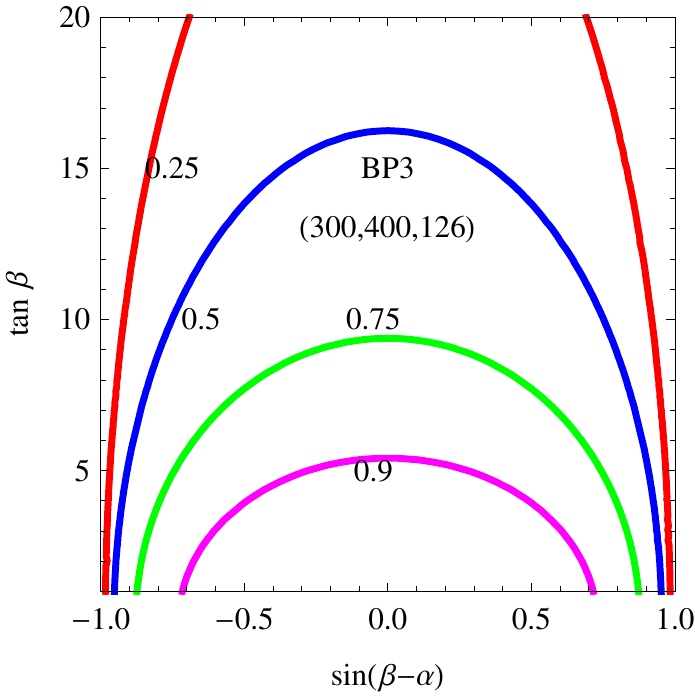}
\caption{Contour plot of $\textrm{BR}(\A\to \h Z)$ for BP1 (left panel), and BP3 (right panel).   Also marked in each plot is   the corresponding values of $(m_{\A},\,m_{\H},\,m_{\h})$ for each benchmark point.}
\label{fig:Sigma1A}
\end{figure}

In Fig.~\ref{fig:Sigma1A}, we show contour plots of BR($\A\to\h Z$) for BP1 (left panel) and BP3 (right panel).  BR($\A\to\h Z$)  always maximizes at $\sba=0$, and  decreases for larger $|\sba|$,   since  $g_{Z\A\h }\sim\cba$.   For BP1 with $(m_{\A},m_{\H},m_{\h}) =  (400, 126, 50)$ GeV,  both $\A\to\h Z$ and $\A\to \H Z$ open, with the coupling of the latter process  proportional to $\sba$.  Therefore,   BR($\A\to\h Z$) decreases more rapidly when $|\sba|$ gets bigger.  BR($\A\to\h Z$)  decreases at large $\tan\beta$ as $\A \rightarrow bb$ becomes more and more important.     For $m_A > 2 m_t$,  $A \rightarrow tt$ becomes competitive at low $\tan\beta$, which correspondingly reduces BR($\A\to\h Z$) further in that region.    For BP2 with $(m_{\A},m_{\H},m_{\h}) =  (400, 200, 126)$ GeV, the behavior of BR($\A\to\h Z$)  is very similar to that of BP1.    

For BP3 with $(m_{\A},m_{\H},m_{\h}) =  (300, 400, 126)$ GeV, only $\A\to\h Z$ opens with no competitive process from  $\A\to\H Z$ and  $\A\to tt$.  Therefore,  comparing to BP1,  BR($\A\to\h Z$) decreases much slower as    $\sba$ approaches $\pm 1$.   BR($\A\to\h Z$) is also  maximized at smaller $\tan\beta$ due to both the absence of $\A \rightarrow tt$ and the suppression of $\A \rightarrow bb$.   
 
\begin{figure}[h!]
\centering
	\includegraphics[]{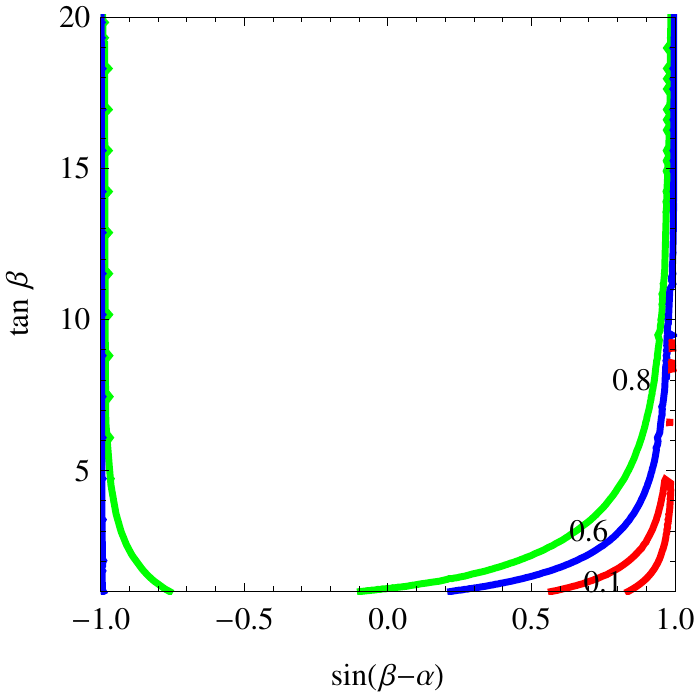}
\caption{Branching ratio of  $\h \rightarrow bb$ for $\h$ being the 126 GeV Higgs.   }
\label{fig:BRhbb}
\end{figure}

To compare with the exclusion and discovery limits in the $bb\ell\ell$, $\tau\tau\ell\ell$  channels, it is also important to know the branching fractions of $\h \rightarrow bb, \tau\tau$, which depend mostly on $m_{\h}$.   For BP1 with $m_{\h}=$ 50 GeV, we used BR($\h \rightarrow bb$)= 82\% and BR($\h \rightarrow \tau\tau$)= 8\%.   For the other benchmark points with $\h$ being the SM-like 126 GeV Higgs,  the branching fraction is obtained by rescaling  the SM value of the BR with relevant coupling coefficients as given in Table.~\ref{tab:couplings}.  We show a contour plot of BR($\h\to bb$)   in Fig.~\ref{fig:BRhbb} for $\h$ being the 126 GeV Higgs.   While $\h \rightarrow bb$ reaches 80\% and saturates in most of the parameter space,  there is a wedge shaped region around $0.5 < \sba < 1$ at small $\tan\beta$ in which $\h \rightarrow bb$ could be suppressed.

\begin{figure}[h!]
\centering
	\includegraphics[width=1.9in]{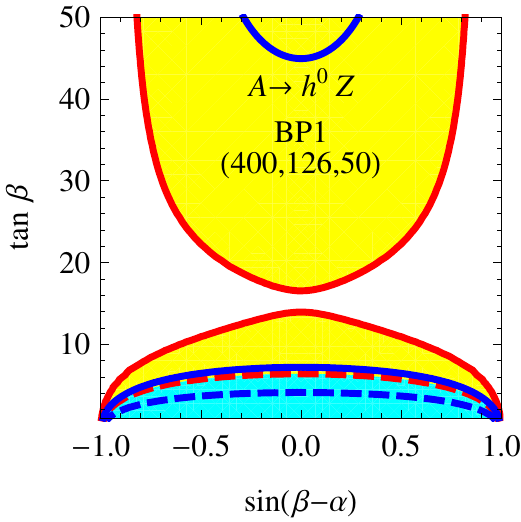}
	\includegraphics[width=1.9in]{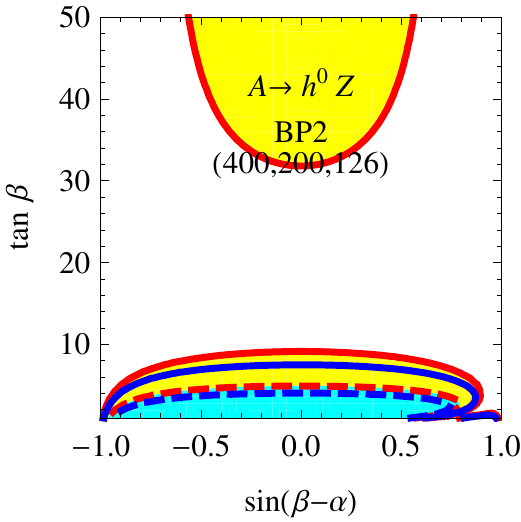}
	\includegraphics[width=1.9in]{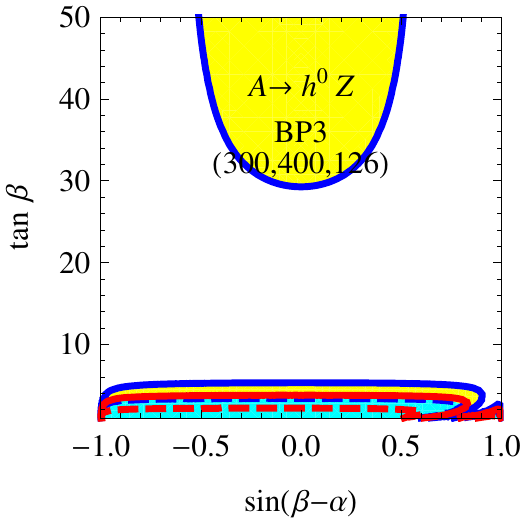}
 \caption{The 95\% exclusion (yellow regions encoded by the solid lines) and 5$\sigma$ discovery (cyan regions enclosed by the dashed lines)  for $gg \rightarrow A \rightarrow \h Z$  in the $\tan\beta$ versus $\sba$  plane, corresponding to an integrated luminosity of 100 fb$^{-1}$ at the 14 TeV LHC for BP1 (left panel), BP2 (middle panel) and BP3 (right panel). The red curves correspond to the $bb\ell\ell$ final state while the blue curves are the results for $\tau\tau\ell\ell$.  Also marked in each plot is the corresponding values of $(m_{\A},\,m_{\H},\,m_{\h})$ for each benchmark point. 
  }
\label{fig:case-1A}
\end{figure}

In  Fig.~\ref{fig:case-1A}, we show the LHC 100 ${\rm fb}^{-1}$ discovery/exclusion reach for $gg \rightarrow \A \rightarrow \h Z$ in the $bb\ell\ell$ (red curves)  and $\tau\tau \ell\ell$ (blue curves) channels for BP1 (left panel), BP2 (middle panel) and BP3 (right panel).   95\% Exclusion regions are shown as yellow regions enclosed by the solid lines while the 5$\sigma$ discovery regions are the cyan regions enclosed  by the dashed lines. Each plot also indicates   the corresponding values of $(m_{\A},\,m_{\H},\,m_{\h})$ for each specific benchmark point.  For all the plots, the discovery region for either case is restricted to $\tan\beta\leq$ 5 where the gluon-fusion cross section is enhanced from the top-loop contribution. For BP1 with $m_A=400$ GeV and a small mass of $m_{\h}=50$ GeV,  the experimental reach on $\sigma\times{\rm BR}$ is the best.    Discovery is possible for all values of $-1 < \sba <  1$ for $\tan\beta$ up to 5, while the exclusion region covers $\tan\beta \lesssim 14$ or large $\tan\beta \gtrsim 16$ with $-0.8 < \sba < 0.8$.   Exclusion or discovery regions with   $\tau\tau\ell\ell$ channel, shown in regions enclosed by the blue curves, are smaller compared to the regions in the $bb\ell\ell$ channel. 
 
For BP2 with $m_A=400$ GeV and $m_{\h}=126$ GeV, regions of $\tan\beta < 10$  or $\tan\beta > 32$ will be excluded if no signal is detected, and regions of $\tan\beta < 4$    can be discovered if there are positive signals.  For BP3 with $m_A=300$ GeV and $m_{\h}=126$ GeV, the exclusion and discovery regions shrink further at small $\tan\beta$.    The wedge-shaped region toward $\sba=$ 1 corresponds to the wedge region in   Fig.~\ref{fig:BRhbb}.   Our results agree with that of Ref.~\cite{Brownson:2013lka} for $\A \rightarrow \h Z$ with $\h$ being the SM-like Higgs.

We note the interesting feature that the $bb\ell\ell$ limits are better than the  $\tau\tau\ell\ell$ ones for BP1 and BP2, while the behavior flipped for BP3.    This is because $\tau\tau\ell\ell$ typically has better reach than $bb\ell\ell$ process at small $m_A$,  while $bb\ell\ell$ does better at large $m_A$,   when the ${\rm BR}(\h\to bb)/{\rm BR}(\h\to\tau\tau)\sim 3 m_{b}^2/m_{\tau}^2$ is taken into account.

Given the smallness of the branching fraction of $\h \rightarrow ZZ$ for the $m_{\h}$ values chosen, the $ZZZ$  channel will  not be useful in probing the parameter space with $gg \rightarrow \A \rightarrow \h Z$.  We also  note that for the $\H$-126 case (BP1) with the favored region to interpret $\H$ as the SM-like Higgs being around $\sba\sim 0$, $gg \rightarrow \A \rightarrow \h Z$ will be extremely useful in probing this region.  For the $\h$-126 case (BP2 and BP3), the favored region to interpret $\h$ as the SM-like Higgs is around $\sba=\pm$1.   Even though the $A\to\h Z$ branching ratio is typically suppressed when $\sba$ approaches $\pm 1$, we could still have reach in $\sba$ extending fairly close to $\pm 1$.  


\begin{figure}[h!]
\centering
	\includegraphics[scale=1.0]{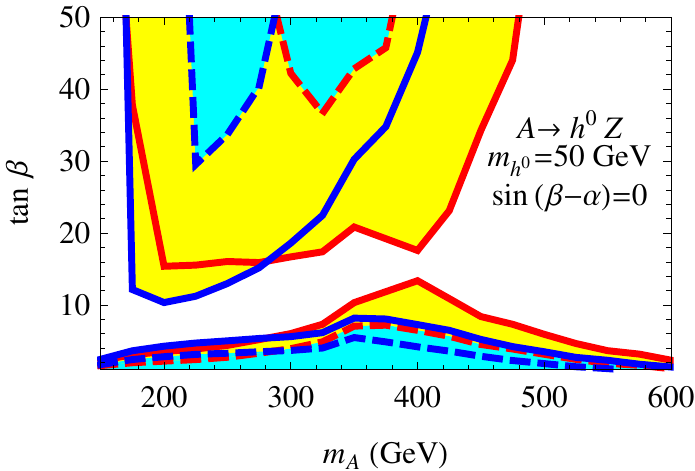}
         \includegraphics[scale=1.0]{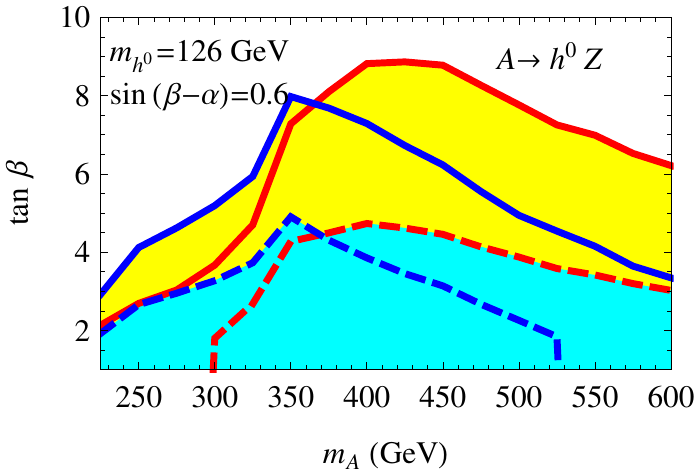}
 \caption{The 95\% exclusion (yellow regions enclosed by the solid curves) and 5$\sigma$  discovery (cyan regions enclosed by dashed curves)   in the $m_{A}-\tan\beta$ plane for $gg \rightarrow A \rightarrow \h Z$ with $m_{\h}=$ 50 GeV, $\sba=$ 0, $m_{\H}=126$ GeV (left panel) and   $m_{\h}=$ 126 GeV, $\sba=$ 0.6, $m_{\H}=$ 1 TeV (right panel), corresponding to an integrated luminosity of 100 fb$^{-1}$ at the 14 TeV LHC.   In either plot, the red and blue curves refer to the limits of $bb\ell\ell$ and $\tau\tau\ell\ell$ channels respectively.   }
\label{fig:limit-bb}
\end{figure}

In the left panel of Fig.~\ref{fig:limit-bb},  we show the reach in $\tan\beta$ versus $m_A$  plane for $m_{\h}=$ 50 GeV and $\sba=0$,  with 95\% C.L. exclusion (yellow regions enclosed by the solid curves) and 5$\sigma$ discovery (cyan regions enclosed by dashed curves) given for $bb\ell\ell$ channel (red lines) and $\tau\tau\ell\ell$ channel (blue lines).   While $\tau\tau\ell\ell$ is more sensitive at low $m_A$, $bb\ell\ell$ extends the reach at large $m_A$.      In general, small $\tan\beta$ (lower region) or large $\tan\beta$ (top region) are within   reach due to the enhancement of the top and bottom Yukawa couplings in those regions.        For small $\tan\beta \sim 1$, almost all values of $m_A$ up to 600 GeV can be covered, with regions of $m_A$ shrink for increasing $\tan\beta$.  At  large $\tan\beta \gtrsim 10$, small $m_A$ can not be approached due to the weakening of the experimental limit, while large $m_A$ can not be approached  due to the decreasing of the signal cross sections.  
 
In the right panel of Fig.~\ref{fig:limit-bb}, we show the reach in $m_A-\tan\beta$ plane for $m_{\h}=$ 126 GeV and $\sba=0.6$.  Note that we have chose a value for $\sba$ that is consistent with the current Higgs search results \cite{Coleppa:2013dya} of a 126 GeV $\h$ while still allowing a sizable branching fraction for $\A \rightarrow \h Z$.   We have decoupled the heavy CP-even Higgs $\H$ so that $\A \rightarrow \H Z$ does not occur.    Given the reduced branching fraction for $A \rightarrow \h Z$, as well as the worse exclusion/discovery limits, the exclusion and discovery regions are smaller, compared to the left panel with $m_{\h}=$ 50 GeV, $\sba=$ 0.  In particular, only regions with $\tan\beta \lesssim 8$ or a small region in $\tan\beta \gtrsim 50$ around $m_A \sim 450$ GeV are viable.

\subsection{$gg\to\A\to\H Z$}

\begin{figure}[h!]
\centering
	\includegraphics[width=2.5in]{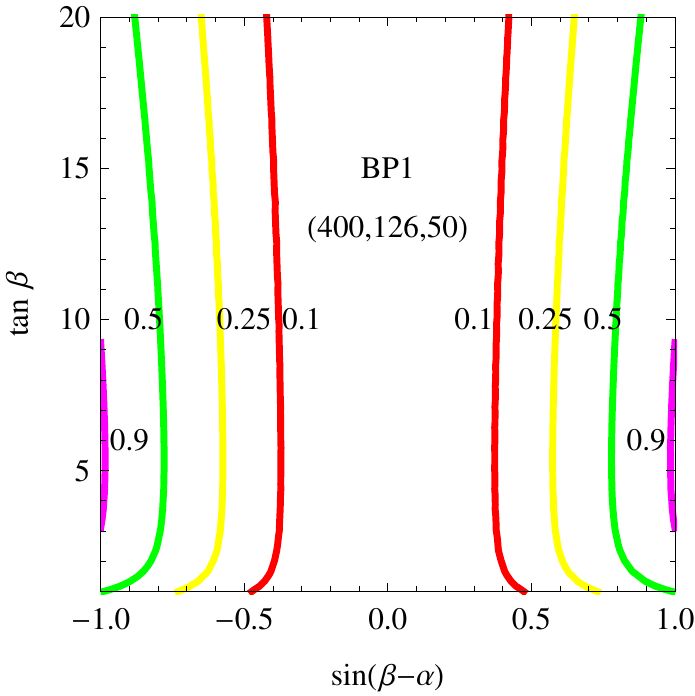}
 	\includegraphics[width=2.5in]{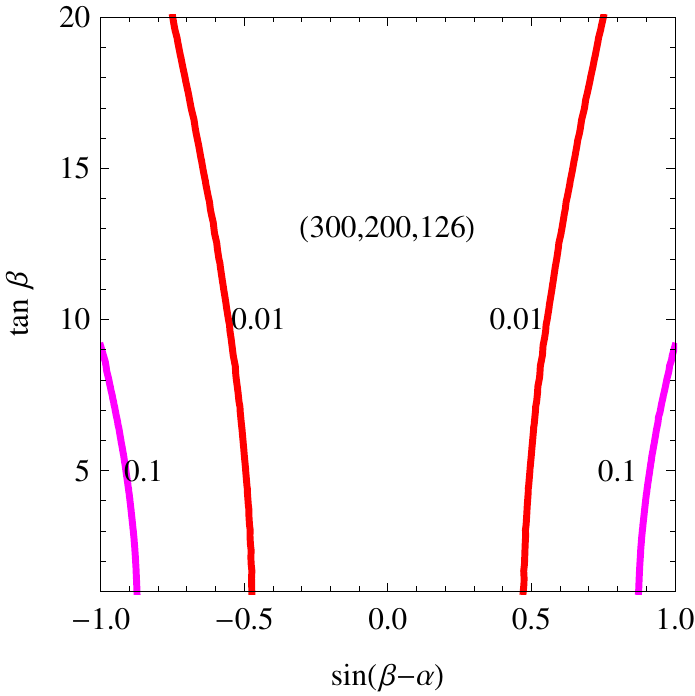}
\caption{Contour plot of $\textrm{BR}(\A\to \H Z)$ for BP1 (left panel) and a comparison point of $(m_{\A},m_{\H},m_{\h})=  (300, 200, 126)$ GeV (right panel).     }
\label{fig:Sigma2A}
\end{figure} 

$A \rightarrow \H Z$ opens once it is kinematically accessible.  Since $m_{\h}<m_{\H}$,  $\A \rightarrow \h Z$ is always accessible and more favorable in phase space.  Whether  $\A\to\H Z$ dominates or not depends largely on $\sba$, which controls the coupling of $Z \A\H $ as well as $Z \A\h $.    Fig.~\ref{fig:Sigma2A} shows the contours of $\textrm{BR}(\A\to \H Z)$ in the parameter space of  $\tan\beta$ versus $\sba$, for BP1 in the left panel. Contrary to the $\A \rightarrow \h Z$ case as shown in Fig.~\ref{fig:Sigma1A},  the   branching ratios become larger for larger $|\sin(\beta-\alpha)|$, which is maximized at $\sba=\pm$1, consistent with Eq.~(\ref{eq:haz-coup}).  While the branching fractions are largely independent of $\tan\beta$, for small $\tan\beta \lesssim 2$,  $\textrm{BR}(\A\to \H Z)$ decreases due to the competition from $A \rightarrow tt$.    The behavior of $\textrm{BR}(\A\to \H Z)$ in BP2 with $(m_{\A},m_{\H},m_{\h}) =  (400, 200, 126)$ GeV is very similar to that of BP1  with  $(m_{\A},m_{\H},m_{\h}) =  (400, 126, 50)$ GeV.  The branching fraction is slightly smaller compared to that of BP1 due to the relatively larger phase space suppression of $\A \rightarrow \H Z$.   As a comparison of the phase space effects,  we show  $\textrm{BR}(\A\to \H Z)$ for $(m_{\A},m_{\H},m_{\h})=  (300,  200, 126)$ GeV in the right panel.  The branching fraction    is  less than 10\% over almost the entire parameter space.   It is also evident that unlike BP1 and BP2, there is no suppression of the branching fractions at small $\tan\beta$ due to the absence of the $tt$ decay mode. 

\begin{figure}[h!]
\centering
	\includegraphics[width=2.5in]{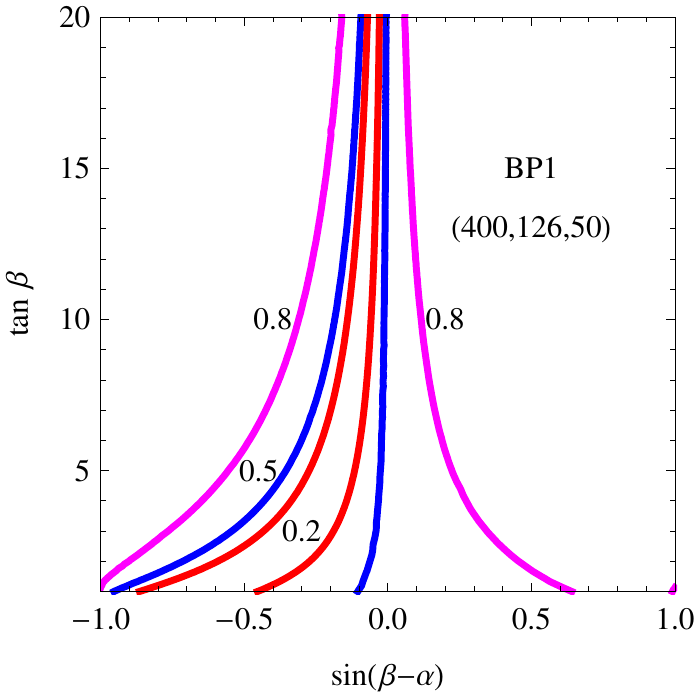}
	\includegraphics[width=2.5in]{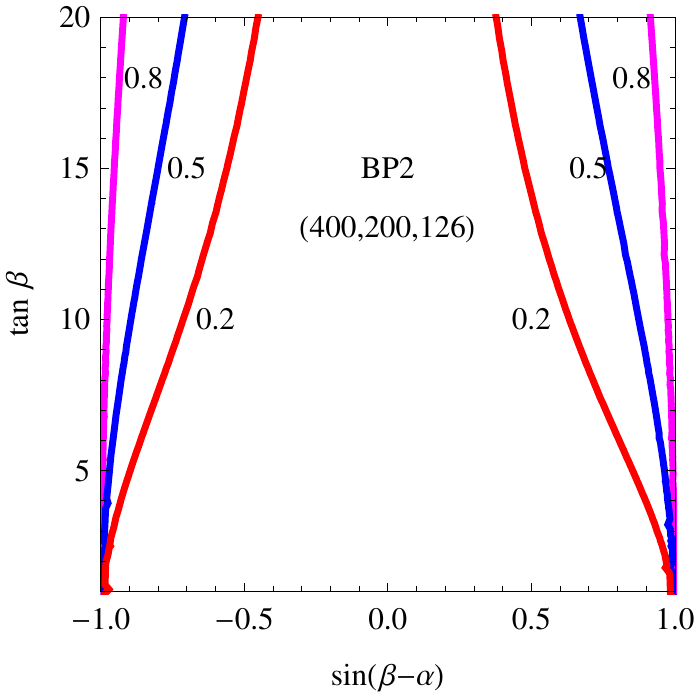}
\caption{Contour plots of $\textrm{BR}(\H\to bb)$ for BP1 (left panel) and BP2 (right panel).   }
\label{fig:BRHbb}
\end{figure}

In Fig.~\ref{fig:BRHbb}, we show contours of the branching ratio $\H\to bb$ for BP1 (left panel) and BP2 (right panel) in  $\tan\beta$ versus  $\sba$ plane.    For BP1 with $m_{\H}=126$ GeV, $\H \rightarrow bb $ is more than 80\%  for $\sba>0.1$ or $\sba<-0.2$ for large $\tan\beta$.  The branching fraction decreases for smaller $\tan\beta$ due to the reduction of the bottom Yukawa coupling.  The further reduction of the branching fraction in negative $\sba$ is due to the  scaling of $\H bb$ coupling as $\cos\alpha/\cos\beta$.     For BP2  with $m_H=200$ GeV, $\H \rightarrow VV$ is kinematically accessible, which reduces $\H\to bb$ further for small $\sba$.    Note that for all the benchmark points chosen, $m_{\H}<2\,m_{t}$, and hence there is no suppression of the $bb$ mode for small $\tan\beta$ when the $tt$ mode would potentially dominate.

\begin{figure}[h!]
\centering
	\includegraphics[width=2.5in]{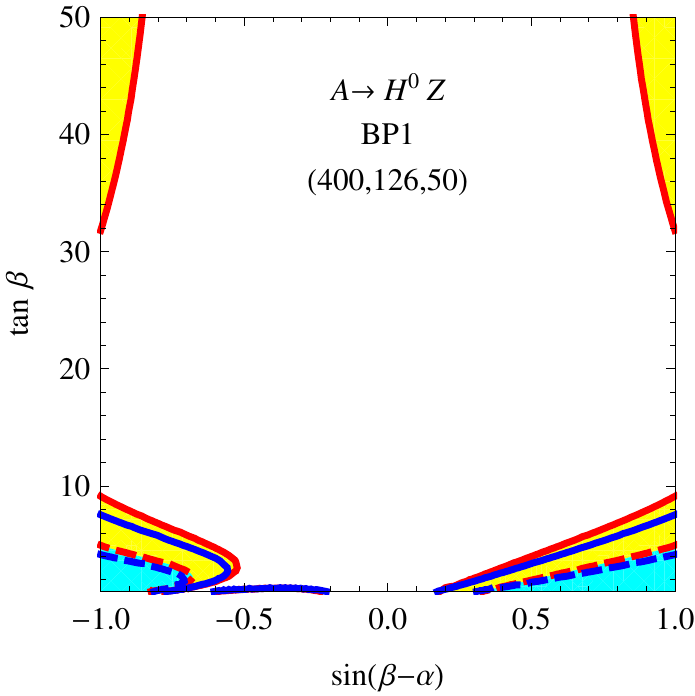}
	\includegraphics[width=2.5in]{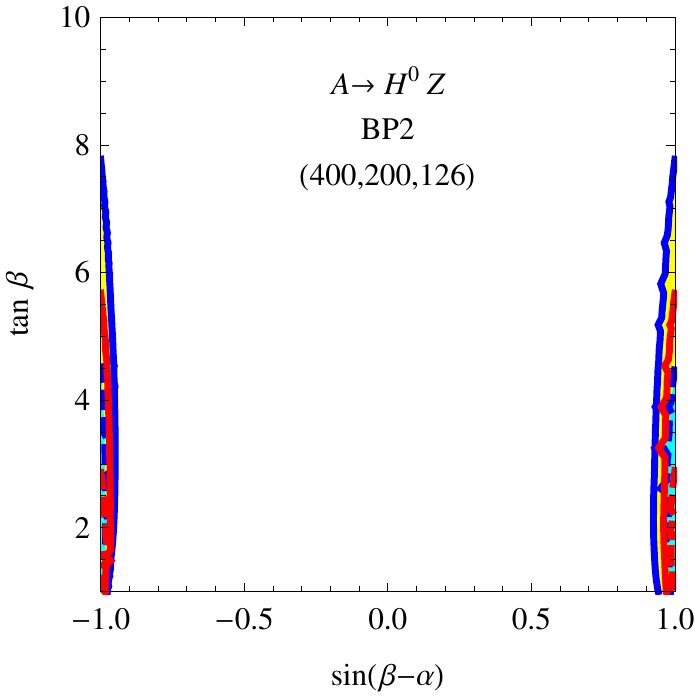}
 \caption{The exclusion  and discovery    region for $gg \rightarrow A \rightarrow \H Z$ in the $bb\ell\ell$ and $\tau\tau\ell\ell$   channels in the $\tan\beta$ versus  $\sba$   plane,  corresponding to an integrated luminosity of 100 fb$^{-1}$ for BP1 (left panel) and BP2 (right panel).   Color coding is the same as in Fig.~\ref{fig:case-1A}.}
\label{fig:Hbb}
\end{figure}

Fig.~\ref{fig:Hbb} shows the  exclusion reach (yellow regions enclosed by the solid lines)  and discovery (cyan region enclosed by the dashed lines)  of $\A \rightarrow \H Z$  for both the $bb\ell\ell$ (red) and $\tau\tau\ell\ell$ (blue) channels.    Regions around $\sba \sim \pm 1$ are reachable while regions around $\sba \sim 0$ are inaccessible due to the suppression of $\A \rightarrow \H Z$.     For BP1, $\tan\beta \lesssim 10$ can be excluded while $\tan\beta \lesssim 5$ is discoverable for $\sba = \pm 1$. The bottom loop effect kicks in at $\tan\beta \gtrsim 32$, excluding slices of parameter space around $\sba=\pm$1.   For $\tan\beta \sim 3$, $-1 \lesssim \sba \lesssim -0.5$ can be excluded, while for $\sba >0$, the exclusion reach extends to $\sba \gtrsim 0.2$ for small $\tan\beta$.   There is also a small additional bump around  $\sba=-$0.6, mainly due to the increasing of BR$(\H\to bb)$, as shown in the left panel of Fig.~\ref{fig:BRHbb}.  The reach is greatly reduced for BP2 due to the suppression  of $\H \rightarrow bb$, except for $\sba \sim \pm 1$.  Only thin slices of parameter region near $\sba \sim \pm 1$ can be covered, which extends to  $\tan\beta \lesssim 8$ for the exclusion, and $\tan\beta \lesssim 4.5$ for discovery.     

Note that for BP1 with $(m_{\A},m_{\H},m_{\h}) =  (400, 126, 50)$, both $\A \rightarrow \h Z$ and $\A \rightarrow \H Z$ open.  The former is more sensitive to the $\sba \sim 0$ region, as shown in the left panel of Fig.~\ref{fig:case-1A},  while the latter is more sensitive to $\sba \sim \pm 1$, as shown in the left panel of Fig.~\ref{fig:Hbb}.   Searches in these  two channels are complementary to each other.  When combined, they could cover the entire region of $\sba$, in particular, for $\tan\beta \lesssim 10$.   Note that when combined with the current experimental search results for the 126 GeV Higgs being the $\H$,  the region with $\sba \sim 0$ is favored, with a thin slice of extended region at negative $-0.8 < \sba <-0.05 $ as well \cite{Coleppa:2013dya}.  

Similar complementarity between  $\A \rightarrow \h Z$ and $\A \rightarrow \H Z$ can be found for BP2 with $(m_{\A},m_{\H},m_{\h}) =  (400, 200, 126)$ GeV, for the entire region of $\sba$.  Interpreting $\h$ being the 126 GeV observed Higgs boson, furthermore, favors $\sba \sim \pm 1$ or a thin slice of  extended region at $0.55 \lesssim \sba \lesssim 0.9$  \cite{Coleppa:2013dya}.

\begin{figure}[h!]
 	\includegraphics[scale=1.0]{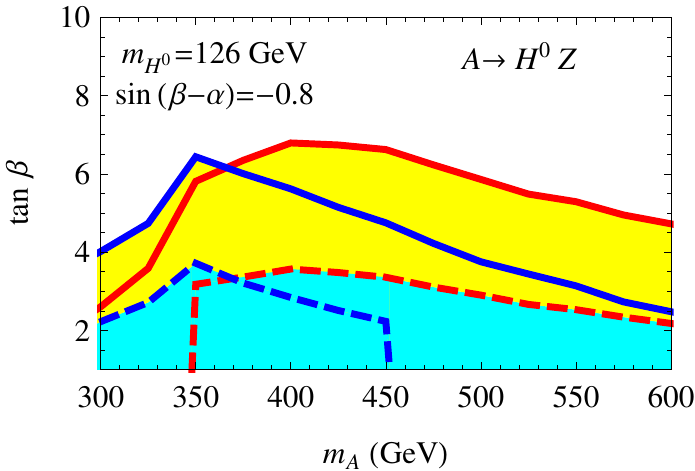}
	\includegraphics[scale=1.0]{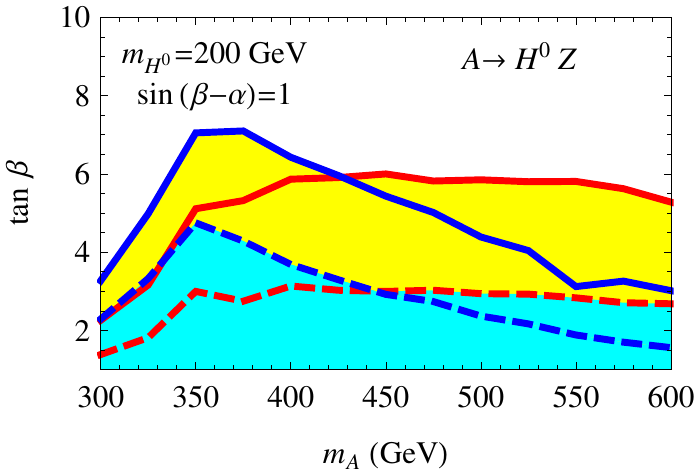}
 \caption{The discovery  and exclusion  regions in the $m_{A}-\tan\beta$ plane for   $gg \rightarrow A \rightarrow \H Z$ in $bb\ell\ell$ and  $\tau\tau\ell\ell$  final states with  $m_{\H}=$ 126 GeV, $m_{\h}=50$ GeV, $\sba=-0.8$ (left panel) and   $m_{\H}=$ 200 GeV, $m_{\h}=$ 126 GeV,  $\sba=$1 (right panel), corresponding to an integrated luminosity of 100 fb$^{-1}$ at the 14 TeV LHC.    Color coding is the same as in Fig.~\ref{fig:limit-bb}.   }  
\label{fig:AHZreach}
\end{figure}

 In  the left panel of Fig.~\ref{fig:AHZreach}, we present the  exclusion  and discovery reach  in the  $\tan\beta$ versus $m_A$ plane for $\A \rightarrow \H Z$ with $m_{\H}=126$ GeV, $m_{\h}=50$ GeV and $\sba =-0.8$.  We have chosen the value of $\sba$ such that the branching faction of $\A \rightarrow \H Z$ is sizable while still consistent with the experimental Higgs search results \cite{Coleppa:2013dya} with a 126 GeV $\H$.   We see that $\tan\beta$ up to about 6.5 can be reached for exclusion, and $\tan\beta$ up to about 3.5 can be reached for discovery.

In  the right panel of Fig.~\ref{fig:AHZreach}, we present the  exclusion  and discovery reach  in the $\tan\beta$ versus $m_A$ plane for   $m_H=200$ GeV, $m_{\h}=126$ GeV and $\sba = 1$.    For 350 GeV $\lesssim m_A \lesssim 600$ GeV, $\tan\beta$ up to about 6 can be excluded, and up to about 3 can be discovered in the $bb\ell\ell$ channel.  $\tau\tau\ell\ell$ channel does better in the low $m_A$ region.

\begin{figure}[h!]
\centering
	\includegraphics[width=2.5in]{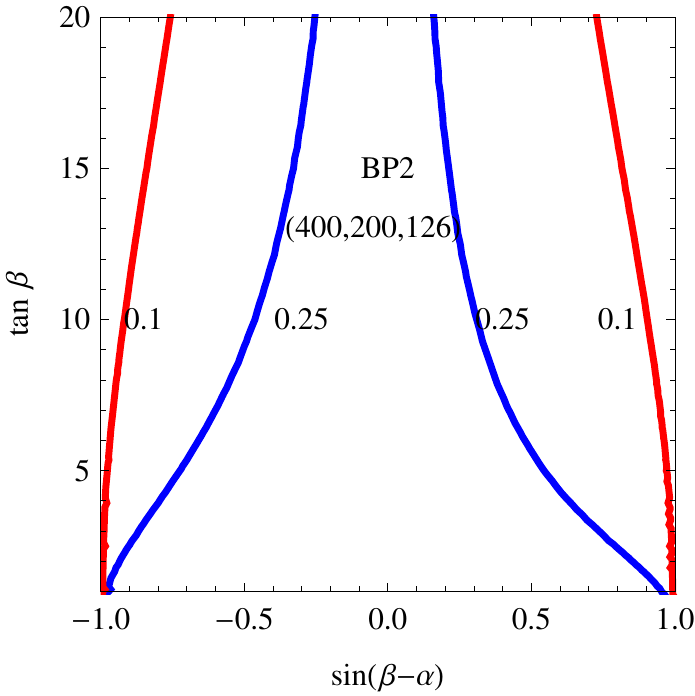}
	\includegraphics[width=2.45in]{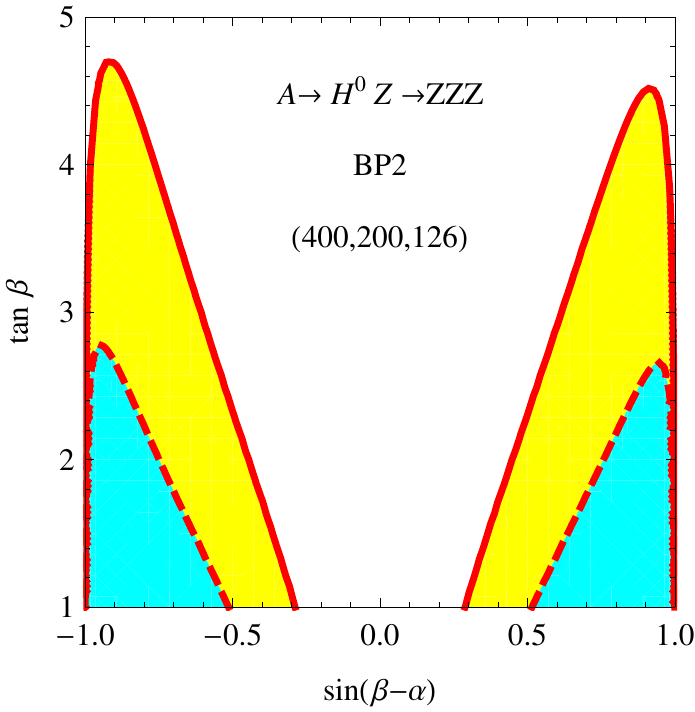}
\caption{Left: Contour plots of $\textrm{BR}(\H\to ZZ)$ for BP2.   Right: The exclusion (yellow regions enclosed by the solid lines) and the discovery (cyan regions enclosed by the dashed lines)  in the $\sba-\tan\beta$ plane for $gg \rightarrow \A \rightarrow \H Z \rightarrow ZZZ$,  corresponding to an integrated luminosity of 100 fb$^{-1}$ at the 14 TeV LHC for the $4\ell+2j$ final state.    }
\label{fig:BR-case2A}
\end{figure}

For BP2 with $m_{\A}=$ 400 GeV, $m_{\H}=$ 200 GeV,   we can also study the parameter reach of $\A \rightarrow \H Z$ with $\H \rightarrow ZZ$.   In Fig.~\ref{fig:BR-case2A}, we show $\textrm{BR}(\H\to ZZ)$ in the left panel, which reaches a maximum of 25\%    for   $|\sba|\lesssim 0.2$.  It gets larger  for small $\tan\beta$ when $\H \rightarrow bb$ is further suppressed.   In the right panel of Fig.~\ref{fig:BR-case2A}, we show the discovery and exclusion contours in the  $\tan\beta$ versus $\sba$ plane for 100 fb$^{-1}$ luminosity  at the LHC.  While $\H \rightarrow ZZ$ maximizes at $\sba \sim 0$, $\A \rightarrow \H Z$ is minimized in this region.  As a result,   regions of  $0.3  \lesssim |\sba| \lesssim 1$  with $\tan\beta$ up to 4.7 can be excluded while the discovery regions are $0.5  \lesssim |\sba| \lesssim 1$ with $\tan\beta \lesssim 2.8$.    Note also that this channel is   complementary to $\A \rightarrow \H Z \rightarrow bb/\tau\tau \ell\ell$ as shown in Fig.~\ref{fig:Hbb}, which is  sensitive to 
$\sba \sim \pm 1$ region.

\begin{figure}[h!]
\centering
	\includegraphics[scale=1.0]{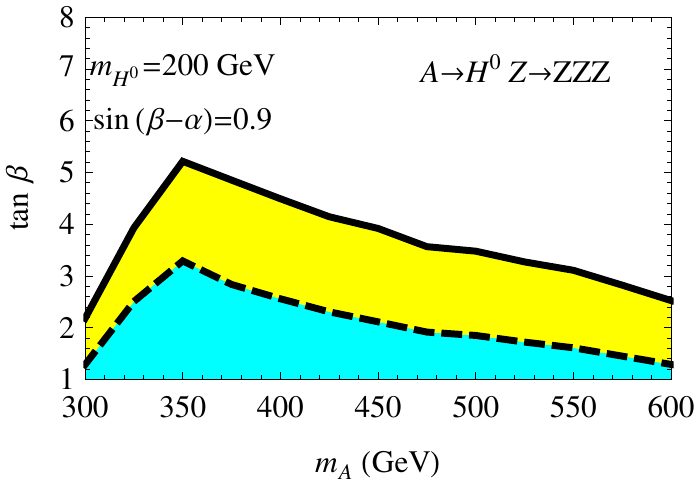}
\caption{The exclusion (yellow region enclosed by the solid lines) and the discovery (cyan region enclosed by the dashed lines)   in the $m_{A}-\tan\beta$ plane with $m_{\H}=$ 200 GeV, $m_{\h}=$ 126 GeV, and $\sba=$ 0.9, corresponding to an integrated luminosity of 100 fb$^{-1}$ at the 14 TeV LHC for the $4\ell+2j$ final state.  }
\label{fig:AHZreach_ZZ}
\end{figure}

In  Fig.~\ref{fig:AHZreach_ZZ}, we present the  exclusion and discovery  in $\tan\beta$ versus $m_A$ plane for $gg \rightarrow \A \rightarrow \H Z \rightarrow ZZZ (4 \ell 2j) $ with $m_H=200$ GeV, $\sba = 0.9$.   We have chosen the value of $\sba$ such that the branching fractions of  both $\A \rightarrow \H Z$ and $\H \rightarrow ZZ$ is sizable while still consistent with the experimental Higgs search results \cite{Coleppa:2013dya} with a 126 GeV $\h$.    We see that the whole region of 300 GeV $< m_A < $ 600 GeV can be covered  at small $\tan\beta$, with the maximum reach in $\tan\beta$   obtained for $m_A \sim$ 350 GeV: $\tan\beta \lesssim 3$ for discovery and $\tan\beta \lesssim 5$  for exclusion.

\subsection{$gg\to\H\to\A Z$}

\begin{figure}[h!]
\centering
 	\includegraphics[width=2.5 in]{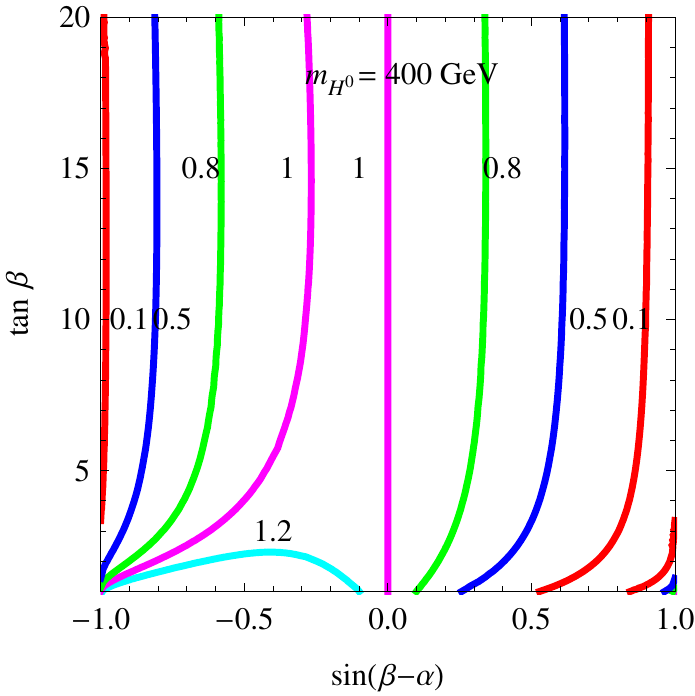}
	\includegraphics[width=2.47 in]{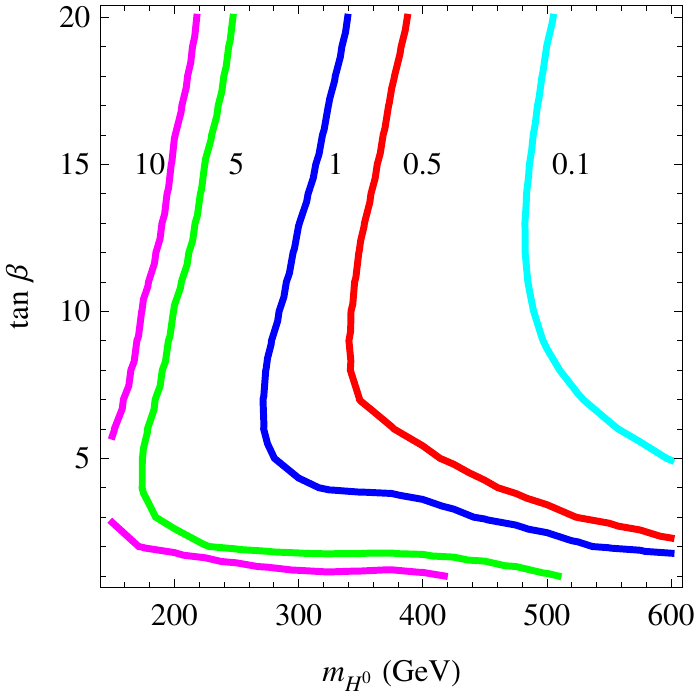}
 \caption{Contours of the cross section normalized to the SM value in the $\sba-\tan\beta$ plane (left panel) for $m_{\H}=400$ GeV  and   $gg\to\H$ cross section at the 14 TeV LHC in unit of pb in the $m_{\H}-\tan\beta$ plane (right panel) with $\sba=-1$.   }
\label{fig:SigmaH}
\end{figure}

For this process, we restrict  to  the $m_{\h}=$ 126 GeV case with a heavier $\H$.   We use BP4 with $(m_{\A},m_{\H},m_{\h})=  (50, 400, 126)$ GeV and  BP5 with $(m_{\A},m_{\H},m_{\h})=  (200, 400, 126)$ GeV as an illustration.  The gluon fusion production cross section for $\H$ can be rescaled from the SM cross section: 
 
\begin{equation}
 \sigma(gg\rightarrow \H)={\sigma_{\rm SM}}\times \frac{\left\vert \left(\frac{\sin\alpha}{\sin\beta}\right) F^{h}_{1/2}(\tau_t)+\left(\frac{\cos\alpha}{\cos\beta}\right) F^{h}_{1/2}(\tau_b)\right\vert^2}{|F^{h}_{1/2}(\tau_t)+F^{h}_{1/2}(\tau_b)|^2},
 \label{eq:Hsigma}
\end{equation}
where the loop factors $F$'s are defined in Eq.~(\ref{eq:loopfactors}).   We note that in contrast to the production of $\A$ in  Eq.~(\ref{eq:sigmaA}), the production of $\H$ involves both $\alpha$ and $\beta$. In the left panel of Fig.~\ref{fig:SigmaH}, we show  contours of  the production cross section of $\H$ normalized to the SM value in the $\sba-\tan\beta$ plane for $m_{\H}=400$ GeV. We see that for positive $\sba$, the cross section is always relatively more suppressed than that for negative $\sba$, introduced by the  interference between the top and bottom loops in Eq.~(\ref{eq:Hsigma}).   For $\sba=\pm 1$, which is preferred by the interpretation of $\h$ being the SM-like Higgs,  the cross section receives the strongest suppression: only 10\% of the corresponding SM value.   In the right panel of Fig.~\ref{fig:SigmaH}, we show contours of the production cross section at 14 TeV LHC in the $m_{\H}-\tan\beta$ plane. We see that cross sections of 10 pb or more is possible for $m_{\H}$ up to 425 GeV for   small $\tan\beta$ -  slightly lower than the corresponding numbers for   $\sigma(gg\to\A)$ as shown in Fig.~\ref{fig:SigmaAfull}. However, the bottom loop enhancement plays a slightly more significant   role in this case at large $\tan\beta$, compared to the $\A$ case.

\begin{figure}[h!]
\centering
       \includegraphics[width=2.5 in]{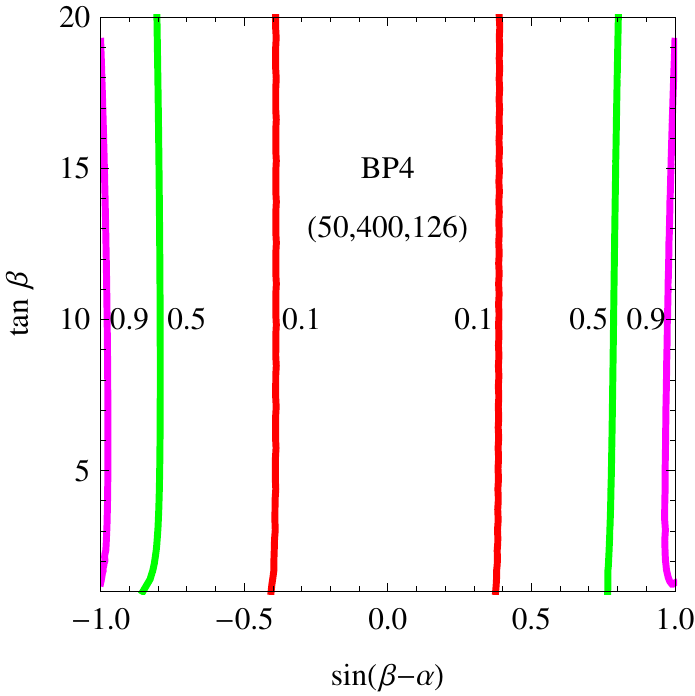}
	\includegraphics[width=2.5 in]{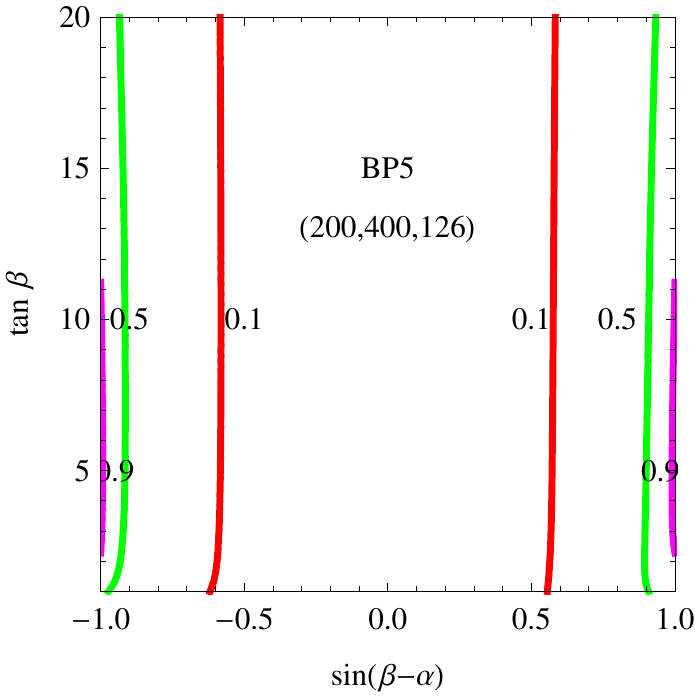}
\caption{Contour plot of  $\textrm{BR}(\H\to \A Z)$ (left panel)  for BP4 (left panel) and BP5 (right panel).  
 }
\label{fig:BR_HtoAZ}
\end{figure}

Fig.~\ref{fig:BR_HtoAZ} shows the $\textrm{BR}(\H\to \A Z)$ for BP4 (left panel) and BP5 (right panel). Since $g_{Z \A\H}\propto\sba$,  the branching fraction gets bigger for larger $|\sba|$, and is maximized at $\sba=\pm$1.  Branching fractions in BP4 is larger than that of BP5 due to the bigger phase space for $\H \rightarrow \A Z$.   For $A\rightarrow bb$ and $\tau\tau$, the branching fraction is about 94\% and 6\% respectively, which does not vary much for BP4 with $m_A=50$ GeV and BP5 with $m_A=200$ GeV.  
 
\begin{figure}[h!]
\centering
	\includegraphics[width=2.5in]{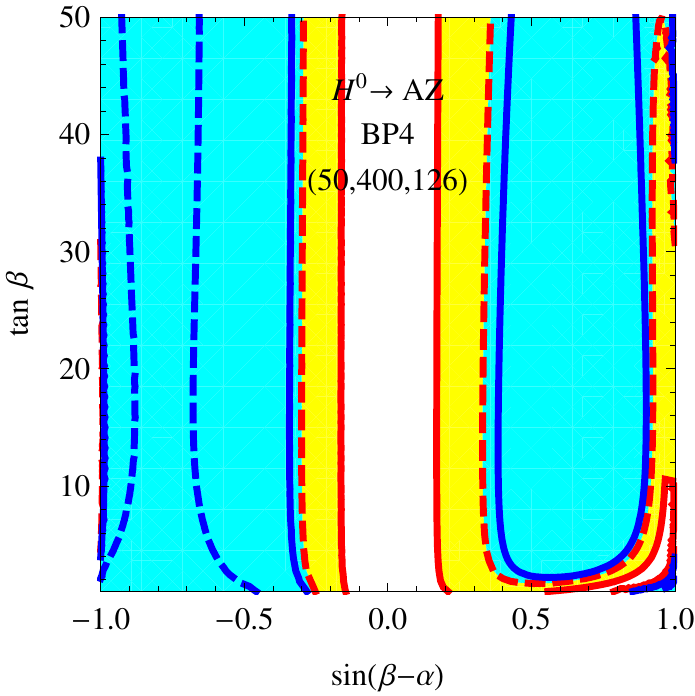}
	\includegraphics[width=2.5in]{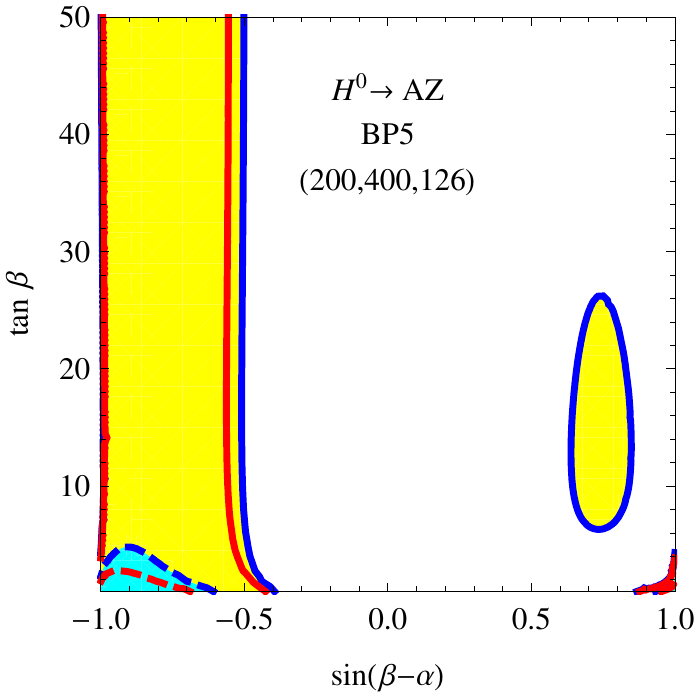}
\caption{The       exclusion and discovery regions   in the  $\tan\beta$ versus $\sba$ plane  for $gg \rightarrow \H \rightarrow \A Z$ with  $bb\ell\ell$ (red) and  $\tau\tau\ell\ell$  (blue) final states for BP4 (left panel) and BP5 (right panel), corresponding to 100 ${\rm fb}^{-1}$ integrated luminosity at the 14 TeV LHC.   Color coding is the same as in Fig.~\ref{fig:case-1A}. }
\label{fig:case-3}
\end{figure}

In   Fig.~\ref{fig:case-3},  we display the discovery/exclusion reach  in $gg \rightarrow \H \rightarrow \A Z$ for the $bb\ell\ell$  (red) and $\tau\tau\ell\ell$ (blue)  final states for BP4 (left panel)  and BP5 (right panel).   For BP4, large regions of parameter spaces in $\tan\beta$ versus $\sba$ can be excluded, except for $-0.15 < \sba < 0.2$ when $\H \rightarrow \A Z$ is highly suppressed.  The discovery region shrinks to $-1 \lesssim \sba \lesssim -0.3$ and $0.35 \lesssim \sba \lesssim 0.9$ for all values of $\tan\beta$.    For BP5, regions  of $-1 \lesssim \sba \lesssim -0.5$ for all $\tan\beta$ and $0.6 \lesssim \sba \lesssim 0.8$ with $6 \lesssim \tan\beta \lesssim 26$ can be excluded and a smaller region in $-1 \lesssim \sba \lesssim -0.6$ with $\tan\beta  \lesssim 5$ can be discovered.  While $bb\ell\ell$ channel has better reach for BP4,  $\tau\tau\ell\ell$ channel has a slightly better sensitivity for BP5.  The reach is also much better for negative $\sba$ because of the less suppressed  cross 
sections of $gg \rightarrow  \H$.

\begin{figure}[h!]
\centering
	\includegraphics[scale=1.0]{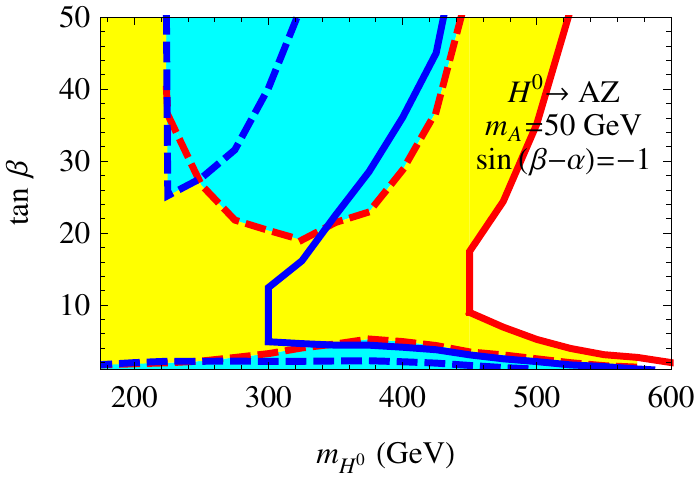}
 	\includegraphics[scale=1.0]{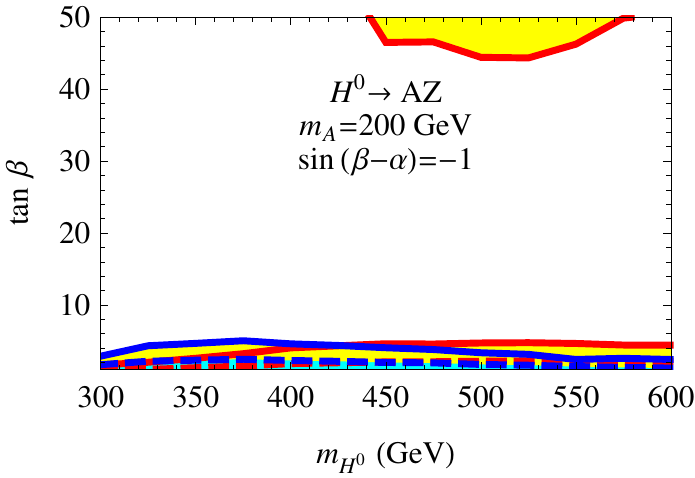}
\caption{ 
The discovery  and exclusion region in the $m_{\H}-\tan\beta$ plane for $gg \rightarrow \H \rightarrow \A Z$ with  $bb\ell\ell$ (red) and  $\tau\tau\ell\ell$  (blue) final states, corresponding to  100 ${\rm fb}^{-1}$ integrated luminosity at the 14 TeV LHC.  The left panel is for $m_A=50$ GeV with $\sba=-1$ and the right panel is for $m_A=200$ GeV with $\sba=-1$.   Color coding is the same as in Fig.~\ref{fig:limit-bb}.    
  } 
\label{fig:limitHAZ}
\end{figure}

In the left panel of Fig.~\ref{fig:limitHAZ}, we show the exclusion and  discovery  each with 100 ${\rm fb}^{-1}$ luminosity at 14 TeV LHC in $\tan\beta$   versus $m_H$   plane,   for $gg \rightarrow \H \rightarrow \A Z$ with  $bb\ell\ell$ (red) and  $\tau\tau\ell\ell$  (blue) final states.  We have chosen $m_A=50$ GeV  and $\sba=-1$.    Discovery is   possible for small values of $\tan\beta \lesssim 5$ or larger values of $\tan\beta \gtrsim 20$.       The exclusion reach, however, is much more extended.   All values of $\tan\beta$ can be covered for $m_{\H}$ up to 450 GeV, with reach extended further at larger and smaller values of $\tan\beta$.   The reach with daughter particle mass $m_A=200$ GeV is shown in the right panel of  Fig.~\ref{fig:limitHAZ}.  Both the exclusion and discovery regions shrink greatly.  Only very small $\tan\beta \lesssim 4$ or very large $\tan\beta \gtrsim  44$ can be excluded.    Note that while $\sba=\pm 1$ is preferred by the interpretation of the $\h$ being the SM-like Higgs,  the suppression of $gg \rightarrow \H$ in that region results in a reduced exclusion/discovery reach.  Even a small deviation of $\sba$ away from $\pm 1$ would introduce a much larger reach in $gg \rightarrow \H \rightarrow \A Z$.  
  

 \section{Conclusion}
\label{sec:conclusions}

Given the discovery of a 126 GeV SM-like Higgs boson at the LHC, it is now time  to use the experimental data   to constrain new physics models while also exploring the detectability of extra Higgs bosons in the extensions of the SM. In this spirit, we explored the production and decay of heavy scalar and pseudoscalar states via the processes  $gg\to \H \rightarrow \A Z$ and $gg\to\A\rightarrow \h Z /\H Z$ with both fermionic ($bb$, $\tau\tau$) and possible  bosonic ($ZZ$) decays of the daughter Higgs.  This channel provides nice complementarity to the conventional search channel   $pp \rightarrow \A/H \rightarrow \tau\tau$,  which is mostly sensitive to the large $\tan\beta$ region.   We presented model independent limits on the 95\% C.L. exclusion and $5\sigma$ discovery in those channels at the 14 TeV LHC. The possibilities   include the interesting case of having the 126 GeV SM-like  Higgs as a decay product of a heavy pseudoscalar.

For the 14 TeV LHC with 300 ${\rm fb}^{-1}$ integrated luminosity, the 95\% C.L. limits on $\sigma \times {\rm BR}$ for the $bb\ell\ell$ final state (where the $b$'s come from the Higgs in the final state) for a 126 GeV daughter Higgs particle vary between  200 fb  to a few fb for the parent heavy Higgs mass in the range of  200 GeV to 600 GeV, while the limit for 5$\sigma$ discovery is about $3-5$  times larger.   For the $\tau\tau\ell\ell$ channel with  the same range of $\A$ mass, the exclusion bounds are around $5-1$  fb and the discovery reach is about 20 fb $-$ 3 fb.    While the $\sigma\times {\rm BR}$ reach in the $\tau\tau\ell\ell$ channel is in general much better than the $bb\ell\ell$ channel,  owing mostly to more suppressed  backgrounds, it is comparable to $bb\ell\ell$ mode once the branching fraction difference between $bb$ and $\tau\tau$ modes are taken into account in a given model.    $gg\to\A\to\H Z\to ZZZ \to 4\ell 2 j$ is useful for heavy Higgses with $m_{\H}> 2 m_Z$.   For $m_{\H}=$ 200  GeV and  $m_{\A}=$ 400 GeV, exclusion in this channel with  300 ${\rm fb}^{-1}$ integrated luminosity requires as little as 1 fb  in $\sigma\times{\rm BR}$ while 5$\sigma$ discovery needs about 3 fb.

We then discussed the implication of the exclusion and discovery bounds of $bb\ell\ell$, $\tau\tau\ell\ell$ and $ZZZ$ channels in the Type II 2HDM,  studying three classes of processes: $gg\to \A \rightarrow \h Z$,  $gg\to \A \rightarrow \H Z$, and  $gg\to \H \rightarrow \A Z$. We find, in general, that there is a significant portion of the  $\tan\beta$ versus $\sba$ plane that allows discovery/exclusion possibilities   in the $bb\ell\ell$ and $\tau\tau\ell\ell$ final states.  $bb\ell\ell$ and $\tau\tau\ell\ell$ have comparable reach, with $\tau\tau\ell\ell$ being slightly better for low parent Higgs masses and $bb\ell\ell$ being better for higher parent Higgs masses.

Specifically, in the channel $gg\to \A \rightarrow \h Z$ when $\H$ is identified as the SM-like Higgs,  95\% exclusion covers most of the  $\tan\beta$ versus $\sba$ plane  for $m_{\A}$ around 400 GeV.    $\tan\beta<5$ can also be covered by 5$\sigma$ discovery.  On the other hand, the exclusion/discovery range is more restricted  when $\h$ is identified as the SM Higgs.   Typically, we find that for $m_{\A}=$ 400 GeV, discovery region lies between $-1<\sba\lesssim 0.8$ and $\tan\beta\leq 5$, while the exclusion region extends to  $\tan\beta \lesssim 10$ or $\gtrsim 30$.   Note also that even though the reach is always maximized at $\sba\sim0$, it extends to larger values of $|\sba|$ close to $\pm 1$ as well.    A wide range of $m_A$ can be covered at low $\tan\beta \lesssim 10$, while high $\tan\beta$ can only be approached for $m_A \lesssim 500$ GeV.

The case where $\A$ decays to $\H Z$  is  complementary  to   $\A \rightarrow \h Z$  in that the discovery and exclusion regions split into two distinct regions around $\sba \sim \pm 1$. We find that in both the $bb\ell\ell$ and $\tau\tau\ell\ell$ channels,  the discovery reach covers $\tan\beta$ up to about 4, while the exclusion region extends to about 7 for $m_A$ up to about 600 GeV.   Moreover, for $m_{\H}\geq$ 200 GeV, this channel also allows for an exclusion reach with $ZZZ$ final states with  $0.3 < |\sba| < 1$, and $\tan\beta$ up to 4.5 for $m_A$ around 400 GeV.  For small values of $\tan\beta$, a wide range of $m_A$ can be covered either by exclusion or discovery.

 In the last class   $gg\to \H \rightarrow \A Z$, we find that discovery/exclusion regions favor the negative $\sba$ regions, largely due to the parameter dependence of gluon fusion production $\sigma(gg\to\H)$.   For $m_{\H}=400$ GeV and $m_A=50$ GeV,  a wide range of  $\tan\beta$  versus $\sba$ space can be covered, except for a small stripe around $-0.15 < \sba < 0.2$.   For  $m_A=200$ GeV, the regions $-1 \lesssim \sba \lesssim  -0.5$ can be excluded for all values of $\tan\beta$, while only a smaller region at low $\tan\beta$ can be discovered.    For $m_A=50$ GeV and  $\sba=-1$, the exclusion reach in $m_H$  can be as large as 450 GeV for  $\tan\beta$ around 10, which extends even further for smaller and larger $\tan\beta$.

While extra Higgs bosons other than the observed 126 GeV SM-like Higgs exist in many extension of the SM, the searches for those Higgses in unconventional decay channels have just started.  Compared to   conventional search channels of $bb$, $\tau\tau$, $WW$, $ZZ$ and $\gamma\gamma$, these exotic decay modes of heavier Higgses decaying into two light Higgses or one Higgs with one gauge boson can be dominant in certain regions of parameter space.  In this paper, we explored $A/H \rightarrow HZ/AZ$ in $bb\ell\ell$, $\tau\tau\ell\ell$ and $ZZZ$ modes.  Other channels, in particular, those involving charged Higgses can be very promising as well \cite{Tong_Su, other_Hpm, Coleppa_Kling_Su}.

\acknowledgments
We thank Nathaniel Craig, Tao Han, Meenakshi Narain, Peter Loch, and John Stupak for helpful discussions.  This work was supported by  the Department of Energy under  Grant~DE-FG02-13ER41976.
 
 \bibliographystyle{JHEP}

\end{document}